\documentclass[%
 aip,
 amsmath,amssymb,
 reprint,%
]{revtex4-1}

\usepackage{graphicx}
\usepackage{dcolumn}
\usepackage{bm}
\usepackage{xcolor}
\usepackage[utf8]{inputenc}
\usepackage[T1]{fontenc}
\usepackage{mathptmx}
\usepackage{etoolbox}
\usepackage{booktabs}
\usepackage{array}

\makeatletter
\def\@email#1#2{%
 \endgroup
 \patchcmd{\titleblock@produce}
  {\frontmatter@RRAPformat}
  {\frontmatter@RRAPformat{\produce@RRAP{*#1\href{mailto:#2}{#2}}}\frontmatter@RRAPformat}
  {}{}
}%
\makeatother
\begin{document}

\preprint{AIP/123-QED}

\title{The Influence of V-Defects, Leakage, and Random Alloy Fluctuations on the Carrier Transport in Red InGaN MQW LEDs}
\author{Huai-Chin Huang\textsuperscript{1}}
\author{Shih-Min Chen\textsuperscript{1}}%
\author{Claude Weisbuch\textsuperscript{3}}
\author{James S. Speck\textsuperscript{2}}
\author{Yuh-Renn Wu\textsuperscript{1*}}
 \email[Author to whom correspondence should be addressed: ]{yrwu@ntu.edu.tw}
\noaffiliation

\affiliation{ Graduate Institute of Photonics and Optoelectronics and Department of Electrical Engineering, National Taiwan University, Taipei 10617, Taiwan
}
\affiliation{ Materials Department, University of California, Santa Barbara, California 93106, USA
}
\affiliation{ Laboratoire de Physique de la Matière Condensée, CNRS, Ecole polytechnique, Institut Polytechnique de Paris, 91120 Palaiseau, France
}
\date{\today}

\begin{abstract}
Red InGaN-based light-emitting diodes (LEDs) exhibit lower internal quantum efficiencies (IQEs) than violet, blue, and green InGaN LEDs due to a reduction in radiative recombination rates relative to non-radiative recombination rates as the Indium composition increases. Additionally, the larger polarization and band offset barriers between high indium content InGaN quantum wells and GaN quantum barriers increase the forward voltage. In blue and green LEDs, random alloy fluctuations and V-defects play a key role in reducing the forward voltage. When V-defects are present, either naturally or intentionally introduced, they create an alternative path for carrier injection into the MQWs through the V-defect sidewalls. This injection mechanism explains the turn-on voltages of green LEDs. However, in InGaN red LEDs, these two phenomena do not reduce the forward voltage as effectively as in blue and green LEDs, and consequently, the computed forward voltage remains significantly higher than the measured one. Furthermore, currents are observed at low voltages before the turn-on voltage (\(V < \hbar\omega/e = 2.0 \, \text{V}\)) of red LEDs. To address this, we introduce dislocation-induced tail states in the modeling, suggesting that leakage current through these states may play a significant role both below and at turn-on voltages. The simulation also indicates that leakage carriers below turn-on accumulate, partially diffuse in the QWs, screen the polarization-induced barrier in the low injection regime, and further reduce the forward voltage. Despite these beneficial effects, a drawback of dislocation-induced tail states is the enhanced nonradiative recombination in the dislocation line region. This study provides a detailed analysis of device injection physics in InGaN QW red LEDs and outlines potential optimization strategies.

\end{abstract}

\maketitle

\section{Introduction}
In recent years, the study of indium gallium nitride (InGaN) materials has gained significant attention for their applications in micro light-emitting diodes ($\mu$-LEDs). The superior efficiency, thermal stability, and emission characteristics of InGaN-based red-green-blue (RGB) micro-LEDs have made them increasingly desirable, especially as display technologies trend toward smaller chip and pixel sizes \cite{microLED3, microLED4, microLED5, microLED2}.

Regarding efficiency, blue InGaN LEDs achieve peak internal quantum efficiency (IQE) exceeding 90\% at a current density of 1-10 A/cm\(^2\) ~\cite{microLED2, blueLED, blueLED2}. However, improving the IQE of red InGaN LEDs has proven particularly challenging, primarily due to the ``green gap'' with the factors contributing to the green gap being even more pronounced in the red spectral region \cite{greengap, experiment}. A key reason for the lower efficiency of high-indium-content InGaN QWs is the longer radiative lifetime caused by the large polarization-induced electric field in the QWs, which results in reduced electron-hole overlap \cite{polarization4} relative to other non-radiative recombination mechanisms. Another challenge is the lower crystal quality of the QWs due to the relatively low growth temperature required for high-indium-content InGaN QWs. Additionally, the wall-plug efficiency (WPE) is affected by the operating voltage, which is primarily increased due to vertical carrier injection being significantly hindered by the large polarization-induced energy barriers at the quantum well/quantum barrier interfaces and the large band offsets \cite{polarization4, polarization1, polarization2, polarization3}. However, experimental results show that the turn-on voltage of red LEDs is close to its theoretical value, indicating the presence of alternative carrier injection mechanisms beyond vertical injection.

\begin{figure}[htb]
\includegraphics[width=\columnwidth]{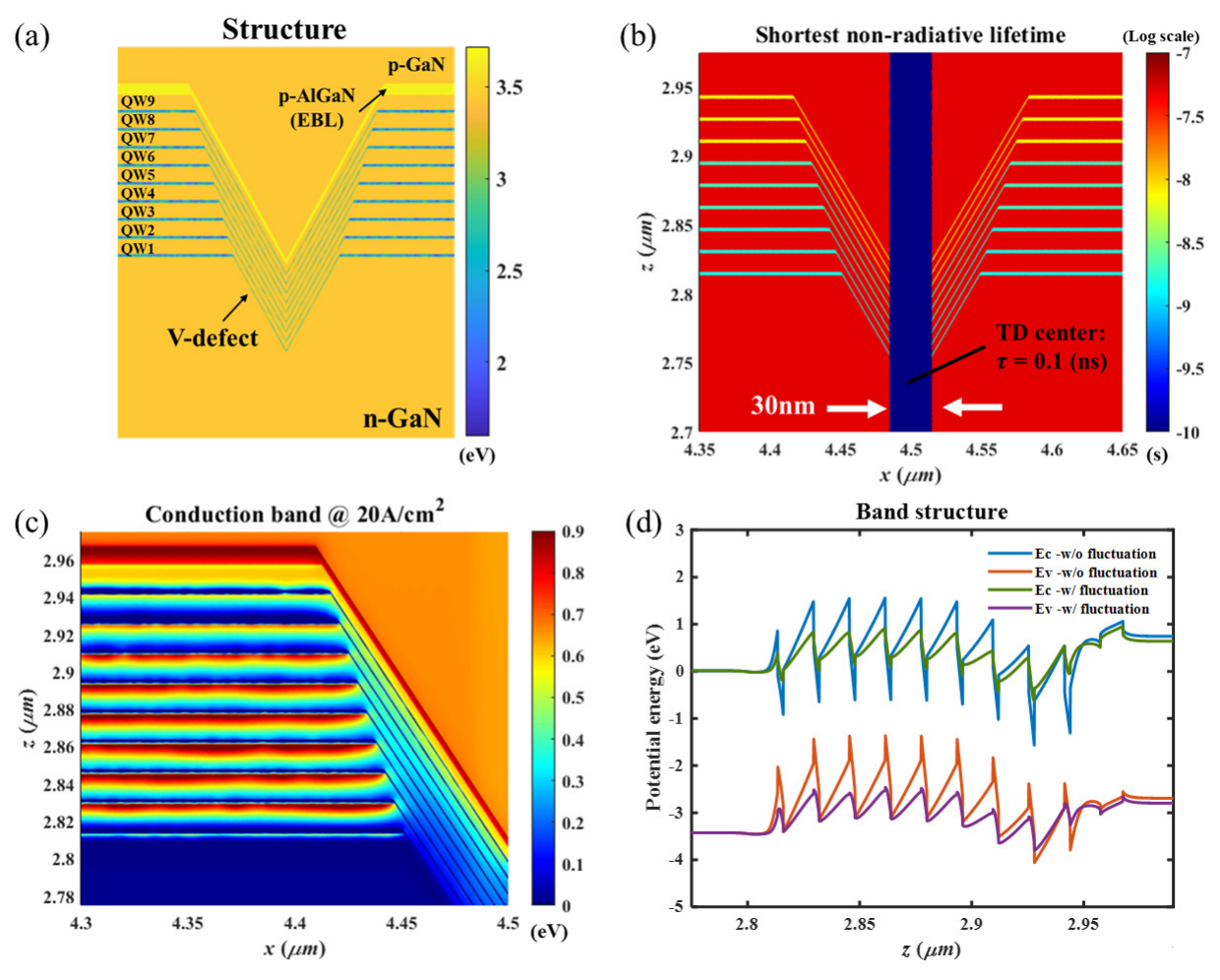}
\caption{(a) Bandgap distribution map of the simulated 9QW Red LED structure with random alloy fluctuations and V-defects. (b) Cross-section of a V-defect and threading dislocation line at the of the V-defect. The TD dislocation line is modeled as a 30 nm diameter cylinder with a high density of fast nonradiative traps, and a short  nonradiative lifetime value in this region. (c) Conduction band potential plotted in log scale. The strong polarization-related and band offset barriers for the c-plane QWs result in carriers flowing through semipolar planes with much-reduced energy variations. (d) Computed band extrema across the vertical LED structure away from the V-defect, taking alloy fluctuations into account (green and violet) or not (blue and red) curves, respectively. In the case of alloy fluctuations, band potential is spatially-averaged over the QW in-plane area. Random fluctuations aid carrier flow by decreasing the effective potential of quantum barriers \cite{LL2}.}
\label{figure_1}
\end{figure}

Two mechanisms that reduce the turn-on voltage in blue and green LEDs have been previously identified and simulated. The natural alloy fluctuation of the InGaN QW material provides local reductions in the barriers for vertical transport in GaN-based blue LEDs (QBs) (Figs. \ref{figure_1}(c) and \ref{figure_1}(d)) \cite{Vdefect1, randomalloyfluctuation1}, enabling percolative paths for currents at lower applied voltages than those required in uniform materials. However, natural alloy disorder alone is insufficient to significantly reduce the barriers for vertical transport in green or red LEDs, leading to their higher forward voltage. One approach to mitigating this issue is carrier injection through V-defects, which are three-dimensional features that form under kinetically limited epitaxial growth conditions (Figs. \ref{figure_1}(a) and \ref{figure_1}(b)) \cite{Vdefect1, Vdefect17, Vdefect11, Vdefect18, Vdefect13, Vdefect16, Vdefect12, Vdefect19, Vdefect14, Vdefect20, Vdefect10, Vdefect15}. Simulations of carrier injection through the shallower and semipolar MQWs of the V-defect sidewalls show that substantial lateral hole injection into the quantum wells occurs at lower forward voltages than for vertical injection, complementing the vertical injection path. Thus, V-defects enhance the LED WPE by reducing the forward voltage \cite{polarization1, Vdefect6, Vdefect8}. However, their presence may also reduce the IQE due to the non-radiative dislocation region of the V-defect \cite{Vdefect8}. In green LEDs, V-defects ultimately improve WPE, as the voltage reduction outweighs the potential decrease in IQE. With sidewall injection from V-defects in green LEDs, current flow becomes lateral in the quantum well, where limited lateral carrier diffusion leads to local current crowding near V-defects and minimal or no improvement in current efficiency droop.

However, in InGaN red LEDs, simulations show that these two phenomena do not reduce the forward voltage to the same extent as in blue and green LEDs. Consequently, the computed forward voltage remains significantly higher than the measured one. Additionally, large leakage currents are observed at low voltages before the turn-on voltage (\(V < \hbar\omega/e = 2.0 \, \text{V}\)) of red LEDs \cite{leakage6, leakage}. To address this, we introduce dislocation-induced tail states in the modeling, as leakage current through such states may play a significant role both below and at turn-on voltages. The simulation further reveals that leakage carriers below turn-on accumulate in the QWs, screen the piezoelectric field-induced barrier in the low injection regime, and further lower the forward voltage. While this effect is beneficial, a drawback of these dislocation-induced tail states is the enhanced nonradiative recombination in the dislocation line region. Our model incorporates tail states in a two-dimensional drift-diffusion charge control (2D-DDCC) simulation, where we successively analyze the effects of: (1) random alloy fluctuations, (2) V-defects, and (3) tail states \cite{randomalloyfluctuation3} on device performance. Computational results indicate that: (1) alloy fluctuations promote radiative and Auger-Meitner recombination due to increased local carrier densities, with the Auger rate increasing more than the radiative rate, leading to a decrease in IQE; (2) V-defects reduce the turn-on voltage by enabling carrier transport across the lower In-content, semipolar sidewall QWs; and (3) carrier transport through tail states near dislocation lines and V-defects provides a better fit to experimental J-V curves at low biases (\(V < \hbar\omega/e = 2.0 \, \text{V}\)) \cite{blueLED, QCSE} and results in a lower turn-on voltage \cite{TD3}.

In this paper, we systematically investigate the impact of V-defects, random alloy fluctuations, and tail states on red InGaN \(\mu\)-LEDs. The influence of carrier injection and the reduction of \(V_{\text{for}}\) will be examined at a current density of 20 A/cm\(^2\). This comprehensive study aims to enhance the understanding of these mechanisms and guide the development of reliable, high-efficiency red InGaN \(\mu\)-LEDs suited for next-generation display technologies.

\begin{figure}
\includegraphics[width=\columnwidth]{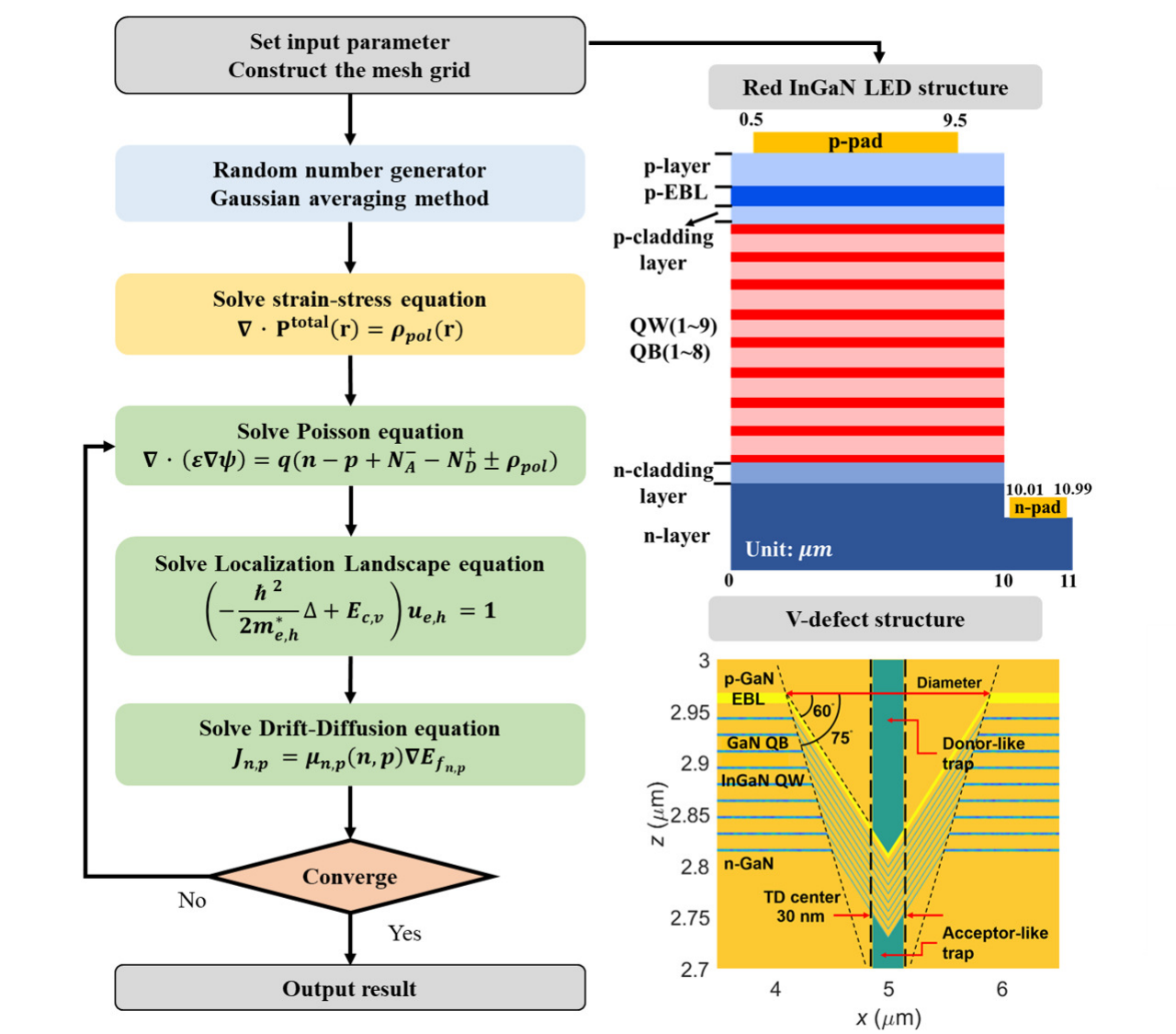}
\caption{Simulation flow chart of the 2D-DDCC program. The parameters of the red InGaN LED structure and the V-defect mesh setting are put in the simulation program.}
\label{figure_2}
\end{figure}

\section{Methodology}
For this research, we employed an in-house two-dimensional drift-diffusion charge-control (2D-DDCC) solver to model carrier transport and the electrical properties of \(\mu\)-LEDs (Fig. \ref{figure_2}) \cite{randomalloyfluctuation2}. This solver is particularly well-suited for analyzing V-defects and random alloy fluctuations, both of which are crucial for understanding device behavior. First, the Poisson equation \eqref{eq1} was used to calculate the electrostatic potential (\(\psi\)) based on the charge distribution. Charge sources include free electrons (\(n\)) and holes (\(p\)), as well as ionized acceptors and donors, \(N_A^{-}\) and \(N_D^{+}\), respectively, while \(\rho_{\text{pol}}\) accounts for the polarization charge density, which depends on the local indium composition. Furthermore, the solver integrates the Localization Landscape (LL) theory \cite{LL2, LL1, LL3}, an advanced method that accounts for both quantum confinement effects and alloy disorder in determining the effective quantum potential \(\frac{1}{u_{e,h}}\). This is achieved by substituting \eqref{eq2} for the conventional Schrödinger equation \cite{LL1}. The model integrates in the energy space using the Fermi-Dirac distribution (\(f_n, f_p\)) and the local density of states for electrons (\(N_c\)) and holes (\(N_v\)) in Eqs. \eqref{eq3} and \eqref{eq4} to compute carrier concentrations, where \(E_{fn}\) and \(E_{fp}\) denote the quasi-Fermi levels for electrons and holes, respectively.

\begin{equation}
    \nabla \cdot (\varepsilon \nabla \psi) = q(n - p + N_A^{-} - N_D^{+} \pm \rho_{\text{pol}})
    \label{eq1}
\end{equation}

\begin{equation}
    \left( \frac{\hbar^2}{2m^*_{e,h}} \Delta + E_{c,\nu} \right) u_{e,h} = 1
    \label{eq2}
\end{equation}

\begin{equation}
    n = \int_{1/u_e}^{+\infty} N_c(E) \cdot f_n(E) \, dE
    \label{eq3}
\end{equation}

\begin{equation}
    p = \int_{-\infty}^{1/u_h} N_v(E) \cdot f_p(E) \, dE
    \label{eq4}
\end{equation}

To obtain the quasi-Fermi levels \(E_{fn}\) and \(E_{fp}\), the drift-diffusion solvers [Eqs. \eqref{eq5} and \eqref{eq6}] need to be solved first.

\begin{equation}
    J_n = - q n \mu_n \nabla E_{fn}
    \label{eq5}
\end{equation}

\begin{equation}
    J_p = - q p \mu_p \nabla E_{fp}
    \label{eq6}
\end{equation}

where \(\mu_n\) and \(\mu_p\) are electron and hole mobilities, respectively. To solve the drift-diffusion equation, we need to consider the equation of continuity to make this equation into 2\(^{\text{nd}}\) order partial different equations as shown in Eq. \eqref{eq7}.

\begin{equation}
    \nabla J_{n,p} = - q (R-G)
    \label{eq7}
\end{equation}

\begin{equation}
    R = R_{\text{SRH}} + B_0 n p + C_0 (n^2 p + n p^2)
    \label{eq8}
\end{equation}

\begin{equation}
    R_{\text{SRH}} = \frac{np - n_i^2}{\tau_{n0}(p + n_i) + \tau_{p0}(n + n_i)}
    \label{eq9}
\end{equation}

Here, \(G\) is the generation rate, and \(R\) is the total recombination rate. \(R_{SRH}\) represents Shockley-Read-Hall (SRH) nonradiative recombination. \(n_i\) is the intrinsic carrier density. The carrier recombination processes, including SRH nonradiative recombination, radiative recombination, and Auger recombination, are modeled using Eqs. \eqref{eq8} and \eqref{eq9}. \(\tau_{n0}\) and \(\tau_{p0}\) are the nonradiative lifetimes of electrons and holes, respectively. \(B_0\) is the radiative recombination coefficient, and the electron-hole overlap in \(np\) accounts for the spatial dependence of the carrier concentrations, i.e., \(n(r)\) and \(p(r)\). \(C_0\) is the Auger recombination coefficient, where the electron-hole overlap is included in \(n^2p\) and \(np^2\). These recombination mechanisms are essential for determining the IQE, as shown in Eq. \eqref{eq10}. Here, we define IQE by integrating carrier recombination in QWs and dividing it by the total injected current, which includes the charge injection efficiency \(\eta_{ie}\). This definition slightly differs from the traditional one. In this paper, carrier overflow is not observed, and \(\eta_{ie}\) is equal to 1.0. The overall WPE, considering IQE, light extraction efficiency (LEE), and photon energy, is given in Eq. \eqref{eq11}.

\begin{equation}
    IQE = \frac{\int_{}^{}q B_0 n p}{\int_{}^{}q R} \times \eta_{ie} = \frac{\int_{}^{}q B_0 n p}{\int_{}^{}q R} \times \frac{\int{}^{}q R}{J} = \frac{\int_{}^{}q B_0 n p}{J}
    \label{eq10}
\end{equation}

\begin{equation}
    WPE = \frac{IQE \times LEE \times \hbar \omega}{qV}
    \label{eq11}
\end{equation}

\begin{figure*}[htb]
\includegraphics[width=14cm]{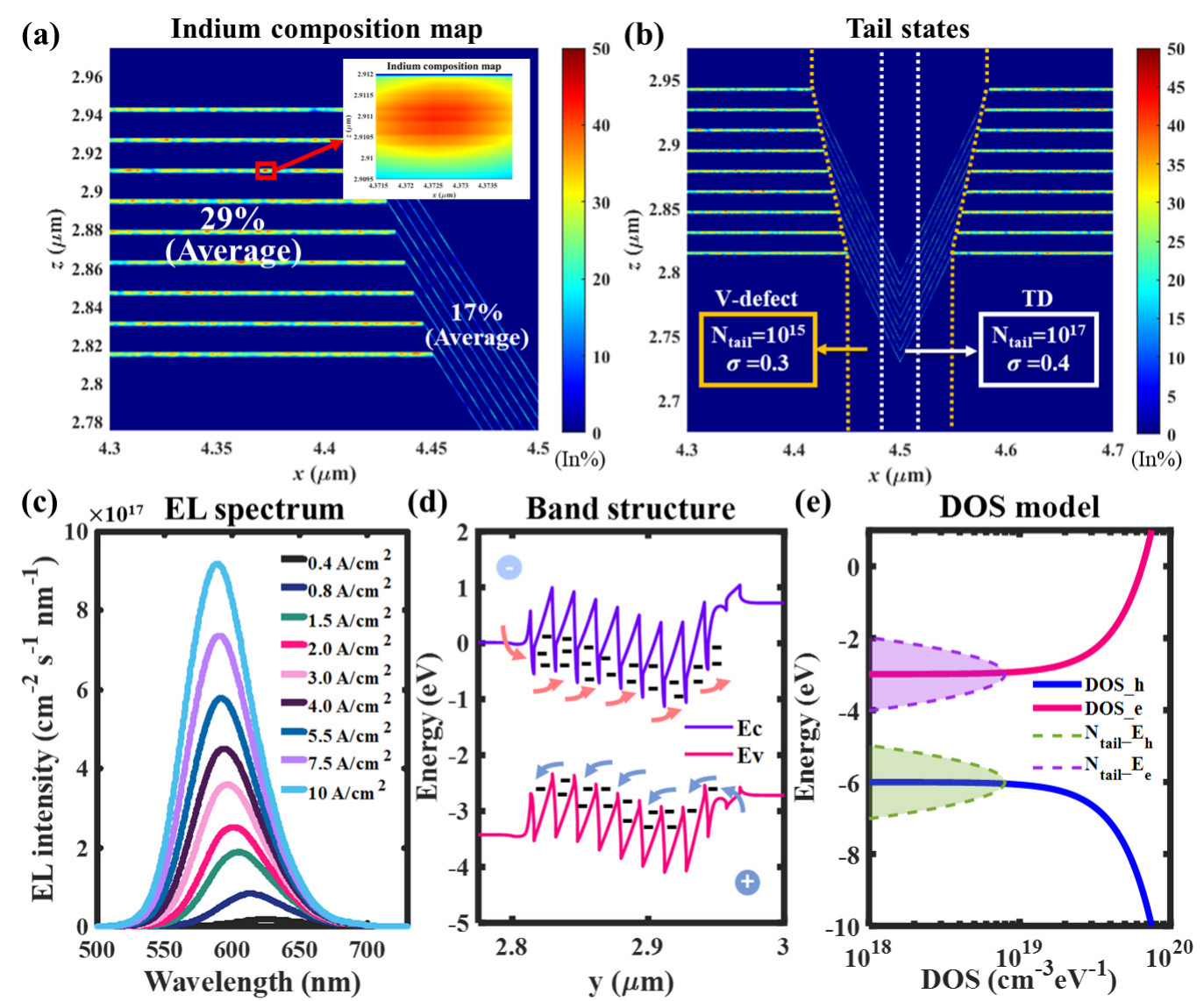}
\caption{Simulation models: (a) Compositional map of the combined model incorporating alloy fluctuations and V-defects. Random alloy fluctuations are illustrated in the inset. Indium composition is 17\% at V-defect sidewalls QWs, averaged to 29\% at c-plane MQWs, and (b) tail states along TD centers and sidewalls, acting as transport and recombination pathways, reducing the turn-on voltage and radiative efficiency, inducing a leakage current below turn-on.  (c) Calculated emission spectrum at different current density with the assigned 29\% random alloy fluctuation. (d) Band structure showing carrier transport through tail states at low current density, leading to an improved carrier injection at low voltage, which is identified as carrier leakage. (e) Position of tail states in the density of states across the bandgap.}
\label{figure_3}
\end{figure*}

To simulate random alloy fluctuations, we used a random number generator combined with Gaussian weighting methods to distribute indium (In) and gallium (Ga) atoms across the lattice sites within the multiple quantum wells (MQWs). This arrangement effectively simulates local compositional variations and their impact on essential metrics such as bandgap energy and polarization, ensuring accurate device characterization. Our model also accounts for the impact of tail states, which are carrier states below the bandgap that arise from crystal structure imperfections, strain-induced bandgap changes, or impurities. These tail-state effects have been widely discussed in polycrystalline, doped, or amorphous semiconductor materials. Such states provide additional carrier percolation paths, significantly influencing carrier dynamics. The origin of tail states in this paper will be discussed later. To incorporate tail states, we define the density of tail states, \(N_{tail}(E)\), represented by a Gaussian distribution [Eq. \eqref{eq12}], as shown in Fig. \ref{figure_3}(e). This equation describes the peak density of tail states \(N_{tail}\), the energy center \(E_{tail}\), and the standard deviation \(\sigma_{tail}\).

\begin{equation}
    N_{(c,\nu),\text{tail}}(E) = N_{\text{tail}} \exp \left( - \frac{(E - E_{\text{tail}})^2}{2 \sigma_{\text{tail}}^2} \right)
    \label{eq12}
\end{equation}

When the Gaussian shape density of states allowing carriers transport in these states is considered in the Poisson drift-diffusion solver, the total density of state for carrier density is modified to

\begin{equation}
    N_{(c,\nu),\text{total}}(E) = N_{(c,\nu)}(E) + N_{(c,\nu),\text{tail}}(E)
    \label{eq13}
\end{equation}

Here, \(N_{(c,v),total}(E)\) replaces \(N_{(c,v)}\) in Eqs. \eqref{eq3} and \eqref{eq4}, allowing carriers to exist in these tail states and hop through them inside the energy bandgap, as illustrated in Fig. \ref{figure_3}(e). Tail states have long been studied in disordered or doped semiconductors. Specifically, tail states have been identified and computed in doped GaAs, where they originate from the local electric fields of ionized impurities \cite{tail8}. These tail-state models have been successfully implemented in our software to describe carrier transport in organic and classical semiconductors \cite{tail4, tail1, tail3, tail2}. Similar functions are now applied here to describe carrier transport in tail states near the TD region.

The simulation framework provides a comprehensive analysis while managing computational complexity. It is limited to a two-dimensional model with a lateral LED chip size of 10 \(\mu\)m. This approach captures the critical effects of tail states, V-defects, and random alloy fluctuations on \(\mu\)-LED performance, offering valuable insights for the development of high-efficiency devices.

\begin{table*}[htb]
\caption{\label{table1} Parameter setting with the red InGaN LED structure.} The structure is grown from the bottom n-GaN layer upwards, with each layer having a specific role, as detailed in Table 3.1.
\begin{ruledtabular}
\begin{tabular}{l c c c c c c c c}
&Thickness&Bandgap&Doping&\(E_a\)&\(B_0\) coefficient&\(C_0\) coefficient&\(e^-\)/\(h^+\) mobility&\(\tau_{n,p}\)\\
Material&(nm) & (eV) & $ \left( 10^{\text{18}} \text{ cm}^{-3} \right) $ & (meV) & $ \left( 10^{\text{-11}} \text{ cm}^{3}/\text{s} \right) $ & $ \left( 10^{\text{-30}} \text{ cm}^{6}/\text{s} \right) $ & $ \left( \text{ cm}^{2}/\text{Vs} \right) $ & $ (\text{ns}) $ \\
\hline
\(p\) GaN & 130 & 3.44 & 30 & 180 & 3 & 0.2 & 100/5 & 50\\
\(p\) AlGaN EBL & 10 & 3.72 & 20 & 264 & 3 & 0.2 & 100/5 & 50\\
\(p\) GaN cladding layer & 13.5 & 3.44 & 20 & 180 & 3 & 0.2 & 100/5 & 50\\
\(n\) GaN QB(1$\sim$8) & 13.5 & 3.44 & 1 & 25 & 3 & 0.2 & 350/10 & 50\\
\(n\) InGaN QW(7$\sim$9) & 2.5 & 2.21 & 1 & 25 & 3 & 1.8 & 150/10 & 7\\
\(n\) InGaN QW(4$\sim$6) & 2.5 & 2.21 & 1 & 25 & 3 & 1.8 & 150/10 & 2\\
\(n\) InGaN QW(1$\sim$3) & 2.5 & 2.21 & 1 & 25 & 3 & 1.8 & 150/10 & 1.5\\
\(n\) GaN cladding layer & 13.5 & 3.44 & 40 & 25 & 3 & 0.2 & 200/10 & 50\\
\(n\) GaN & 2800 & 3.44 & 5 & 25 & 3 & 0.2 & 200/10 & 50\\
\end{tabular}
\end{ruledtabular}
\end{table*}

\subsection{Parameter setting}

The red InGaN LED structure follows the design used in Ref.~[\textnormal{\cite{experiment}}]. The structure is grown from the bottom n-GaN layer upwards, with each layer serving a specific function, as detailed in Table \ref{table1}. The structure begins with a 2800 nm thick n-GaN layer, which acts as the n-type contact layer and facilitates current injection. Above this, the device consists of alternating layers of n-InGaN quantum wells (QWs) and n-GaN quantum barriers (QBs), forming the active region. The QWs, with a thickness of 2.5 nm, are designed for efficient electron-hole recombination, emitting red light. The 13.5 nm thick QBs separate each QW, enhancing carrier confinement and minimizing leakage.

Above the active region, a 10 nm p-Al\(_\text{0.15}\)Ga\(_\text{0.85}\)N electron blocking layer (EBL) is included, as in Ref.~[\textnormal{\cite{experiment}}]. The use of an EBL is common in violet and blue LEDs to prevent electron overflow at high injection currents. However, for longer-wavelength green and red LEDs, this is generally unnecessary due to the deeper QWs, which strongly suppress carrier overflow, as well as the lower operating voltages. As long as the bias voltage remains significantly below the GaN p-n junction built-in voltage, electron overflow is negligible. Here, we retain the EBL to simulate the exact structure from Ref.~[\textnormal{\cite{experiment}}]. The top p-GaN layer, with a thickness of 130 nm, serves as the p-type contact layer, completing the device structure.

\begin{table}[htb]
\caption{\label{table2} Parameter of trap levels located in the TD center. The energy levels of the traps (\(E_t\) are given relative and below the conduction band (\(E_c\)).}
\begin{ruledtabular}
\begin{tabular}{l c c c}
&Density of traps&\(\tau_n,\tau_p\)&\(E_t\) below \(E_c\)\\
Type& $ \left( 1/\text{cm}^{3} \right) $ & (ns) &  (eV)\\
\hline
Donor-like trap & $ 1\times 10^{18} $ & 0.5 & 0.6\\
Acceptor-like trap & $ 1\times 10^{18} $ & 0.1 & 3.0\\
\end{tabular}
\end{ruledtabular}
\end{table}

\begin{table}[htb]
\caption{\label{table3} The parameters for tail states density and FWHM of Gaussian distribution in threading dislocation lines and V-defect sidewalls. \(E_{tail}\) is zero in the setting. The value used here gives a satisfactory fit to the WPE curve shown in Ref.~[\textnormal{\cite{experiment}}]. The QW region has a random alloy fluctuation, and the $\sigma_{tail}$ changes from 0.4eV to 0.3eV for indium composition varies from 0\% to 100\%, respectively. 
}
\begin{ruledtabular}
\begin{tabular}{l c c }
& Tail states            & FWHM of the \\ 
& density (\(N_{tail}\)) & Gaussian distribution \\
 & (\(\text{cm}^{-3}\))  & ($\sigma_{tail}$) (eV) \\

\hline
Threading dislocation line & \(10^{17}\) & 0.4\\
V-defect sidewall QB & \(10^{15}\) & 0.3\\
QW region  &  \(10^{16}\) &  0.4-0.3
\end{tabular}
\end{ruledtabular}
\end{table}

Further material parameters, such as bandgap energies, doping concentrations, and recombination coefficients, are provided in Table \ref{table1}. To account for spatially varying material quality, non-radiative lifetimes ($\tau_{nr}$) were assigned to three distinct Quantum Well (QW) groups: QWs 1-3, QWs 4-6, and QWs 7-9. This physically motivated simplification was adopted because: (1) QWs grown at different stages may exhibit varying defect densities; (2) carrier recombination predominantly occurs in QWs near the p-side (QWs 7-9), making their ($\tau_{nr}$)  most critical; and (3) this approach avoids an excessive number of free parameters and overfitting, unlike assigning individual lifetimes to all nine QWs, while still capturing essential spatial variations in non-radiative recombination. A uniform ($\tau_{nr}$)  model was found insufficient for accurate IQE fitting. The model's sensitivity to these grouped lifetimes is further detailed in the Supplementary Material (Fig. S1-S4). The Auger-Meitner coefficient \(C_0\) is chosen to be significantly larger in the InGaN QW region, as it is well known that alloy disorder substantially increases this coefficient compared to pure GaN \cite{Auger, Auger1}. This choice of \(C_0\) also provides better agreement with the experimental results discussed later.
To accurately model the experimental emission spectrum (peaking around 590-625 nm depending on current density, Ref. [\cite{experiment}]), careful determination of the effective In concentration in the c-plane QWs is crucial, particularly given the TEM-verified QW thickness of ~2.5 nm [Ref. \cite{experiment}]. While Ref. [\cite{experiment}] mentions a nominal ~40\% Indium target, such values are often initial estimates based on Vegard’s law for uniform bulk InGaN bandgaps, which do not fully account for Quantum Confined Stark Effect (QCSE) and random alloy fluctuation in thin QWs. Achieving precise high-In incorporation and sharp interfaces in InGaN QWs is inherently complex, often leading to deviations between nominal targets and actual compositions, as documented in studies on In incorporation challenges and advanced characterization (e.g., Refs. [ \cite{DUBOZ2023127033, 10.1116}]). Furthermore, the presence of random alloy fluctuations means that light emission is often dominated by localized In-rich sites as shown in Fig. \ref{figure_3} (a)). This localization effect contributes to the observed difference between a nominal In target and the effective average composition needed to model the overall spectrum. 

Given these constraints, we systematically investigated different initial local In contents within the 2.5 nm thick c-plane QWs to best match the experimental spectrum. The Schrödinger solver is used to calculate the eigenstates and the emission spectrum shown in Fig. \ref{figure_3}(c) with the fluctuated potential in the c-plane QW. Notably, the bowing coefficient for the InGaN alloy bandgap is taken as -2 eV [\cite{indiumcomposition}]. Our model begins by randomly assigning Indium and Gallium atoms within the QW lattice sites with an initial target local average Indium content (such as 37\% in the final result). To determine the actual local indium composition a Gaussian weighting method is employed [Ref. \cite{LL2}]. This method calculates the local In concentration by applying Gaussian weights to surrounding In and Ga atoms, thereby reflecting the immediate atomic environment. This process inherently considers the influence of the neighboring GaN Quantum Barriers (QBs) where only Ga atoms (zero In) are present near the interface. The resulting spatially varying local Indium composition, exhibiting significant fluctuations, is depicted in Fig. \ref{figure_3}(a) (and conceptually in Supplementary Material, Fig. S5(a)). This provided the optimal fit to the experimental emission spectrum from Ref. [\cite{experiment}] with a peak in-plane average indium composition at 29\%. Crucially, our model, which includes these random alloy fluctuations and the strong QCSE, demonstrates that the actual light emission predominantly originates from localized In-rich sites where the Indium content can significantly exceed this 29\% average, and indeed locally surpass even a 40\% nominal value (as illustrated by the high-Indium regions in Fig. \ref{figure_3}(a)). While the effective average peak for emission calculation is 29\%, the underlying random distribution means local regions with higher Indium concentrations exist and can influence carrier localization and emission spectrum. 

 For the V-defect structure. The thin (0.67 nm) V-defect sidewall QWs were designed with an indium composition of 17\%, along with 3.62 nm thick sidewall QBs. These variations lead to corresponding changes in the bandgap energies and polarization properties of the sidewall QWs, which can impact carrier dynamics and recombination efficiency. In the TD center, within 30 nm regions, there is a high density of deep-level traps and shallow-level tail states. The deep-level trap state parameters are listed in Table \ref{table2}, while the tail state parameters are shown in Table \ref{table3}.

In this study, we include tail states in our model. The possible origins of these tail states are as follows: Near the TD center, crystal distortion occurs, significantly affecting the strain in the surrounding QW/QB regions. In red InGaN QW LEDs, a strain variation of 2–4\%, combined with deformation potentials (\(D_1\) to \(D_6\) \cite{tail7}) ranging from 1.7 to 6.3 eV, can introduce a bandgap variation of several hundred meV. This variation may contribute to the tail states discussed here. Additionally, threading dislocations in GaN generate strong localized electric fields due to their charged nature, primarily through carrier trapping \cite{tail6}. These fields shift the absorption edge to lower energies, broadening the absorption tail and further leading to the presence of tail states. This effect is similar to the edge broadening observed in GaAs, as reported in Ref.~[\textnormal{\cite{tail8}}].

Our findings indicate that tail states significantly influence carrier transport \cite{tail1, tail3}, particularly when strong polarization-induced fields and large band offsets hinder vertical carrier injection into the QWs, as in red LEDs. This disrupts the ideal carrier flow over barriers, as illustrated by the band structure model in Fig. \ref{figure_3}(d). The altered energy landscape redirects carriers, increasing leakage to non-radiative pathways and promoting recombination at NR states near dislocations. Furthermore, the inclusion of tail states (Table \ref{table3}) modifies the density of states (DOS), as shown in Fig. \ref{figure_3}(e). The additional energy levels introduced by tail states increase the probability of carrier trapping, enhancing nonradiative recombination in dislocation regions and reducing radiative efficiency. These effects on the DOS highlight the critical role of tail states in shaping the electronic properties and overall optical performance of the LED.

Considering the equivalent band edge densities of states (DOS) of 2.3 $\times$ \(10^{18}\) cm\(^{-3}\) (\(N_c\)) for the conduction band and 1.8 $\times$ \(10^{19}\) cm\(^{-3}\) (\(N_v\)) for the valence band, along with the random distribution of strain and electric fields caused by dislocations, we selected the values for tail state densities and the full width at half maximum (FWHM) of the Gaussian distribution (\(\sigma_{tail}\)) as presented in Table \ref{table3}. These values were chosen to accurately reflect typical material imperfections and their effects on carrier trapping and recombination processes. When incorporated into our simulations, these parameters closely match the experimental results for internal quantum efficiency (IQE) and current-voltage (I-V) characteristics. A comprehensive analysis of tail state formation will not be explored further in this paper.

\begin{figure}[tb]
\includegraphics[width=\columnwidth]{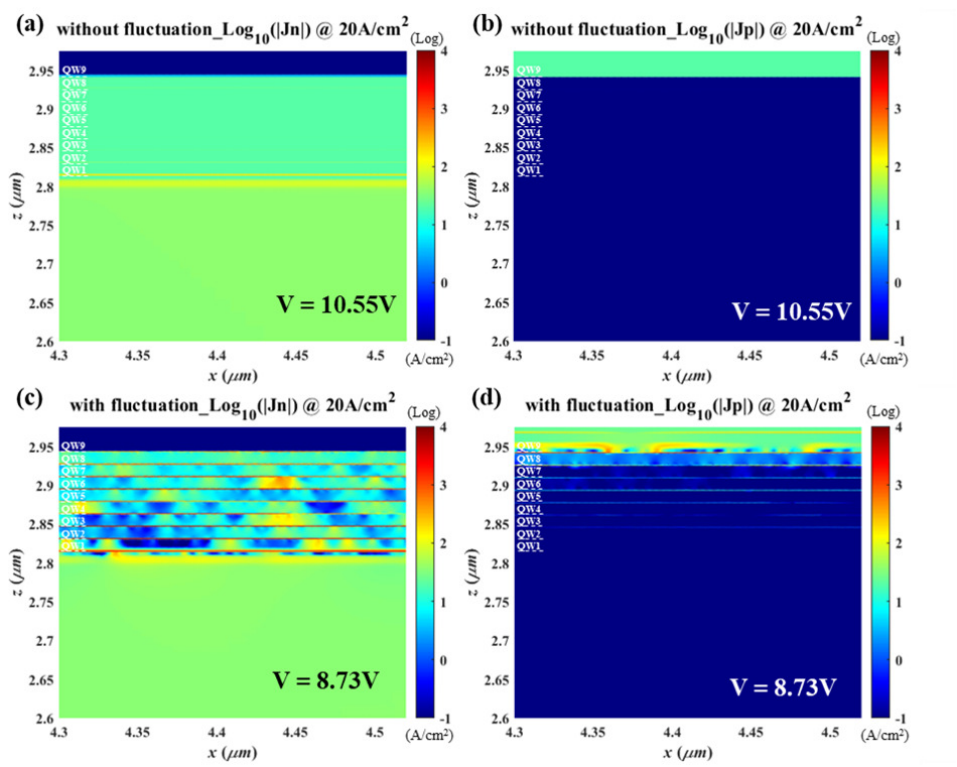}
\caption{The (a, c) electron and (b, d) hole current density distribution of the first two models at current density 20 A/cm\(^2\). In (a) and (b), the applied voltage is V = 10.55V, while in (c) and (d), the applied voltage is V = 8.73V. Fluctuations create fluctuated current flowing paths and higher local carrier concentrations, which may lead to locally enhanced recombination in the higher indium sites.}
\label{figure_4}
\end{figure}

The specific values for parameters such as non-radiative lifetimes and tail state characteristics were determined by optimizing the fit to experimental J-V and IQE data from Ref. [\cite{experiment}], within physically reasonable constraints. While an exhaustive exploration of all parameter combinations is computationally prohibitive, illustrative sensitivity analyses for key parameters (non-radiative lifetimes, different indium compositions for emission peak verification) are provided in the Supplementary Material (Figs. S1-S6) to support our parameterization and validate the model's behavior. To ensure full reproducibility, the final simulation input file for our publicly available 2D-DDCC solver and a corresponding execution guide are also included in the Supplementary Material.

\subsection{Simulation models}
Having defined the material parameters, we simulated four different models to separately analyze the impact of tail states, V-defects, and random alloy fluctuations on the performance of red InGaN LEDs. The first model excluded all three factors. Figure \ref{figure_3}(a) shows the second model, which introduces random alloy fluctuations, followed by the third model, which incorporates V-defects. Finally, the fourth model includes all three factors: random alloy fluctuations, V-defects, and tail states, as shown in Fig. \ref{figure_3}(b). The region containing tail states is highlighted by the yellow dashed line in the V-defect region.

It is important to note that the V-defects incorporated in our simulations (e.g., in the third and fourth models) are not treated as isolated entities. To represent realistic device conditions and investigate density effects, our simulations are performed over a 10 $\mu$m wide domain, typically including multiple V-defects (10 V-defects for baseline results, which corresponds to 10$^{8}$ cm$^{-2}$ density). The similar approach was made in blue and green LED studies\cite{blueLED2}. Accurately resolving the intricate current flow around these multiple defects, alongside random alloy fluctuations, necessitates an extremely fine computational mesh (e.g., dx=0.5nm, dy=0.1nm in active regions), leading to approximately 16.6 million nodes per 2D slice. This results in a substantial computational burden, with each full J-V curve simulation requiring over 80 GB of RAM and 1-5 days of processing time on multi-core CPUs. Despite these demands, this multi-defect approach is crucial for a more accurate assessment of V-defect impact. The influence of varying V-defect densities, achieved by changing the number of explicit V-defects within this domain, is further analyzed later.

It is important to note certain simplifications adopted in our 2D simulation model to maintain computational feasibility while focusing on the key physical phenomena governing the device characteristics under investigation. One such simplification is the omission of the InGaN/GaN superlattice (SL) structure typically present beneath the active region in many experimental LED designs, including that described in Ref. [\cite{experiment}].

Including the full SL stack would lead to a prohibitive increase in computational resources (estimated ~5x memory, significantly extended time), given our already demanding 2D model which incorporates random alloy fluctuations, multiple V-defects, and tail states within a large domain. Our primary focus is elucidating the 2D carrier transport within the MQW active region itself, including V-defect assisted injection and leakage mechanisms. While SLs undoubtedly influence strain and crystal quality, thereby impacting Internal Quantum Efficiency (IQE), their direct effect on the fundamental turn-on voltage characteristics can be less significant when they primarily serve as underlying quality-enhancement or strain-management layers. For instance, studies on GaN-based LEDs incorporating InGaN/GaN SLs underlying the active region have shown significant improvements in crystal quality and optical output, while the turn-on voltage from current-voltage characteristics remained largely unaffected by the presence of such SLs \cite{zhao2020influence,Tao:18}. Tao et. al. shows that more SL even reduce the voltage due to their effect on the V-defect size or densities. This suggests that while crucial for optimizing overall device efficiency through material improvement, the primary electrical injection barriers are still largely dictated by the active MQW, V-defect, and p-n junction properties. Therefore, omitting the detailed SL structure allows us to dedicate computational resources to the complex 2D phenomena within the active region, which are critical for accurately modeling the J-V characteristics and leakage currents. This modeling choice differs from 1D SL simulations [e.g., Ref. \cite{JIANG2019120}] that cannot resolve these critical 2D effects essential for accurate J-V fitting\cite{blueLED2,randomalloyfluctuation2}. This omission is an acknowledged limitation regarding the absolute prediction of IQE values that are dependent on SL-induced quality improvements, but the core electrical transport phenomena are believed to be robustly captured.
\begin{figure*}[tb]
\includegraphics[width=18cm]{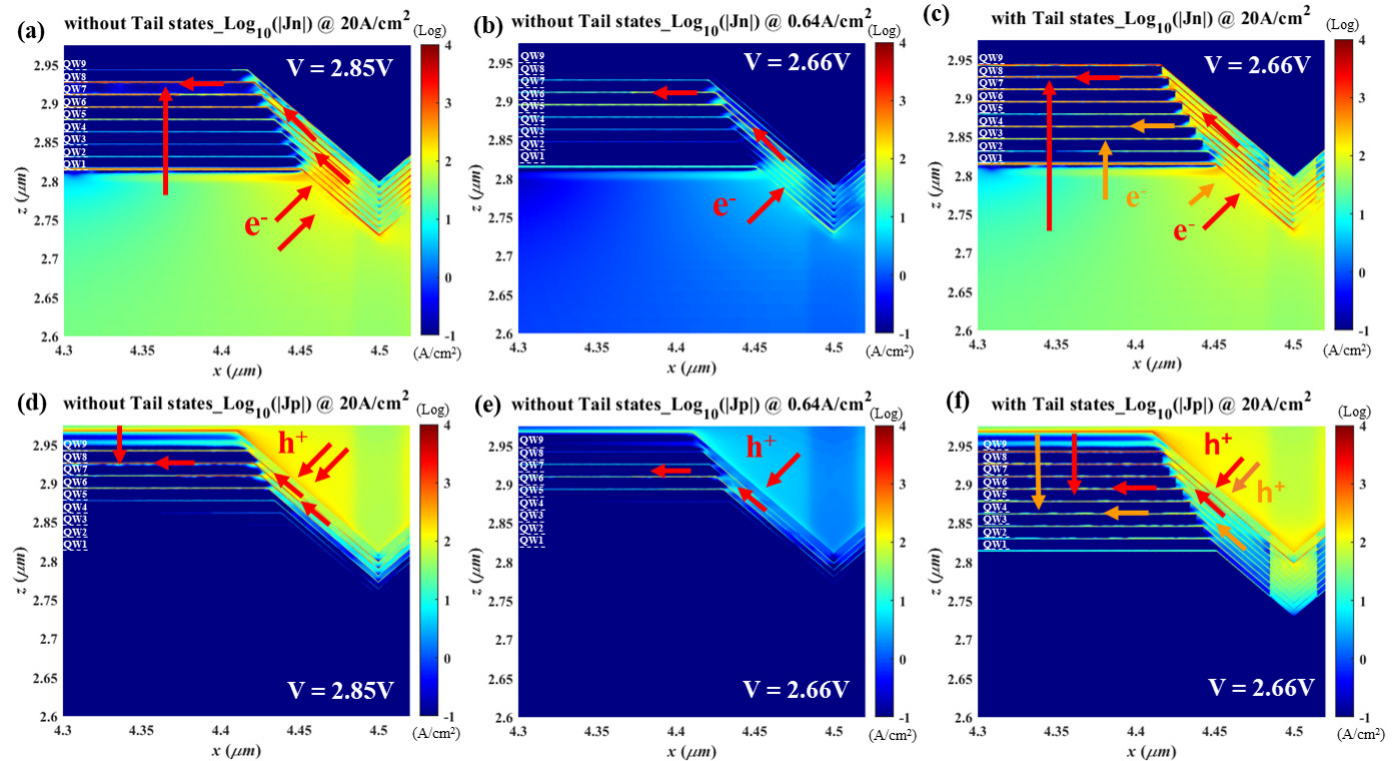}
\caption{Computations of current density distributions of (a, b, c) electrons and (d, e, f) holes at applied voltages of V = 2.85V and 2.66V for the third model (compositional fluctuations + V-defects, (a, b) and (d, e), and for the fourth model (addition of tail states (c, f)).  V-defects improve carrier flow while tail states increase leakage and nonradiative recombination in the defect region. However, holes are then leaking to lower QWs so that more QWs will emit light. Also, a lower \(V_{for}\)  is needed to reach a current density of 20 A/cm\(^2\) compared to the model without tail states.}
\label{figure_5}
\end{figure*}

\section{Results and Discussion}
\subsection{Carrier Injection Mechanism Through V-defects}

In this section, we sequentially examine carrier injection in: (1) homogeneous semiconductor layers, (2) the presence of random alloy fluctuations, (3) the influence of V-defects, and (4) the newly added tail state leakage model. The current flow mechanisms for the four different models are analyzed at a current density of 20 A/cm\(^2\). Figures \ref{figure_4}(a) and \ref{figure_4}(b) illustrate the current flow in the first model, which does not account for random alloy fluctuations, V-defects, or defect state leakage. In this case, current is uniformly injected into the QWs, with carrier injection remaining relatively constant throughout. However, due to the absence of localized states or additional carrier transport pathways, the applied voltage must be very high (10.55V) to reach 20 A/cm\(^2\), indicating significantly hindered vertical carrier injection. As extensively discussed in our previous work on blue and green LEDs \cite{Vdefect16, LL2, randomalloyfluctuation2}, the polarization-induced and band offset barriers between the QWs and QBs are even stronger in red InGaN MQW LEDs. Consequently, the required applied voltage is much larger than in blue and green LEDs, despite the smaller QW bandgap.

In the second model, regions of increased electron and hole flow emerge due to higher local carrier concentrations in the high-indium composition regions of the QWs, introduced via random alloy fluctuations (shown in Figs. \ref{figure_4}(c) and \ref{figure_4}(d)). These compositional fluctuations create potential variations, enabling current flow through lower-barrier sites while also providing localized carrier confinement. Compared to the previous model, the applied voltage (\(\sim\)8.73V) at 20 A/cm\(^2\) is lower than in the uniform composition case, indicating that fluctuations help reduce energy barriers for carrier injection. Nonetheless, the applied voltage remains significantly high, failing to explain or replicate the low operating voltages observed in InGaN red LEDs. When carriers encounter low-energy barrier sites, they preferentially inject through these regions and subsequently diffuse laterally, as illustrated in Fig. \ref{figure_4}(d).

In Fig. \ref{figure_5}(a,d), V-defects and random alloy fluctuations are incorporated into the third model. Since the sidewalls of the QWs are semipolar planes with minimal piezoelectric polarization, the V-defects provide an additional channel for carrier injection. As a result, carrier flow into the c-plane QWs is more efficient through the sidewalls of the V-defects. The reduced barriers, caused by the lower piezoelectric polarization, allow for a more concentrated current density at the V-defect sites, where the voltage at a current density of 20 A/cm\(^2\) significantly drops to 2.85V. Additionally, carriers trapped along the TD lines may lead to nonradiative recombination losses. Although the forward voltage decreases significantly, it remains higher than the experimental value of approximately 2.1 V at 1 A/cm\(^2\). This discrepancy arises because, although the V-defect \(\{ 10\overline{1}1 \}\) semipolar plane (\(\sim 62^\circ\)) exhibits nearly zero polarization at the interface, it typically features a thinner QW with a lower indium composition, similar to that of blue/violet MQWs. The theoretical turn-on voltage of blue LEDs is approximately 2.72V (\(\hbar\omega = 2.72eV\)). Consequently, the sidewall QW aids carrier injection once it becomes conductive. Even if the sidewall QW had a higher indium composition (22\%) instead of the 17\% used, the voltage would decrease slightly but still would not reach the low turn-on voltage due to the increased band offset in the QW and the higher series resistance resulting from the presence of nine deeper sidewall QWs. Therefore, incorporating a tail state current transport model is necessary.

\begin{figure}[tb]
\includegraphics[width=\columnwidth]{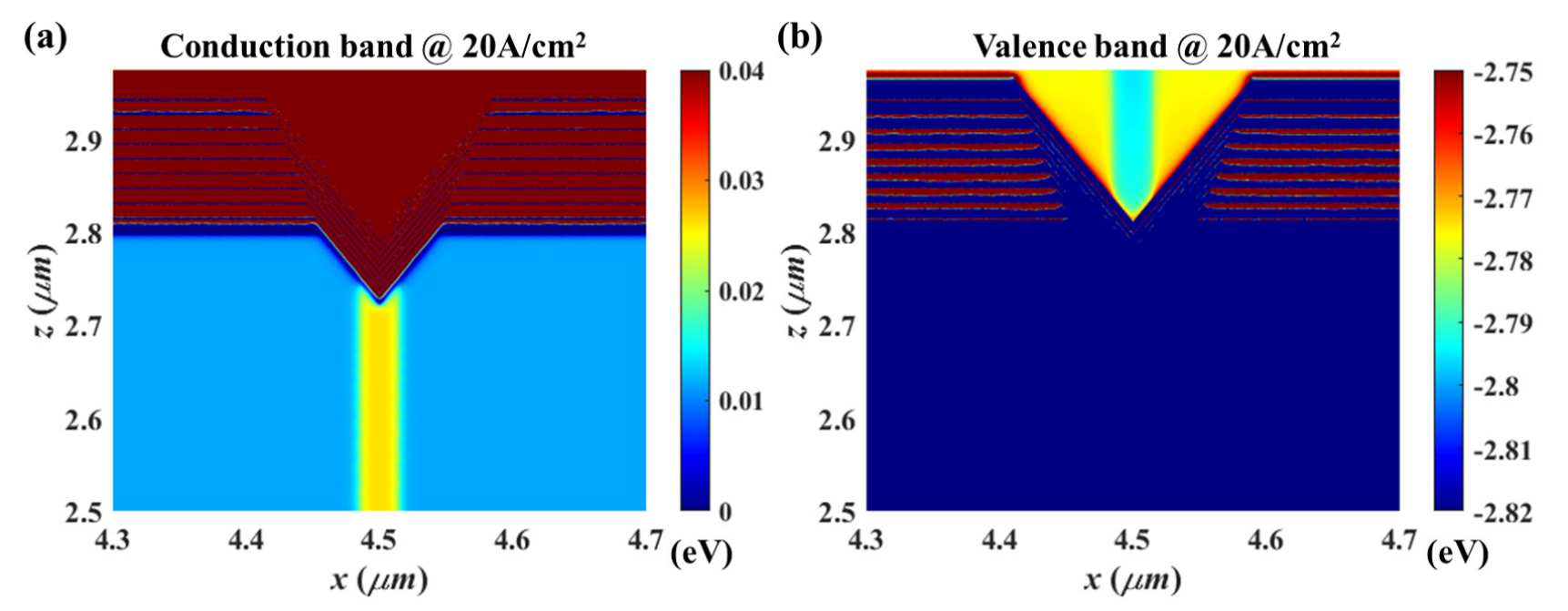}
\caption{The plot of (a) conduction band potential and (b) valence band potential distributions at 20 A/cm\(^2\) with a selected scale range to show the potential barrier induced by electron and hole-trapped charges in the TD center (traps parameters in table \ref{table2}).}
\label{figure_6}
\end{figure}

The interplay between V-defects, random alloy fluctuations, and tail states plays a crucial role in carrier dynamics and device performance. Localized low-energy zones at the TD centers significantly enhance carrier injection, as shown by the conduction and valence band distributions in Fig. \ref{figure_6}. Electrons trapped by deep-level states in the n-layer induce a potential barrier at the TD center, while trapped holes in the p-type layer also create a potential barrier in the TD center. In the active region of the TD, electrons and holes recombine non-radiatively. Furthermore, if tail states are considered, their high density near the TD region may lead to increased local carrier concentrations and currents.

\begin{figure}[tb]
\includegraphics[width=\columnwidth]{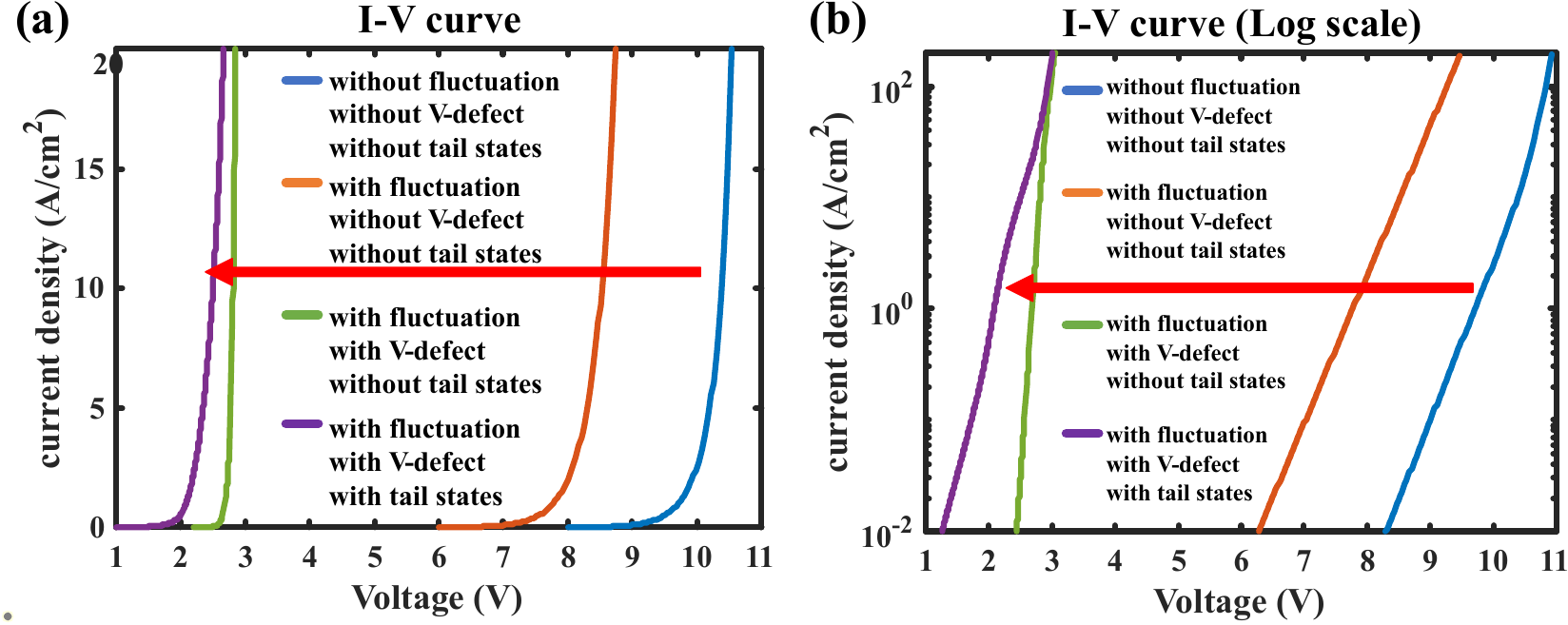}
\caption{(a) The IV curves show the decrease of turn-on voltage as more mechanisms are included, indicating improved carrier injection efficiency. (b) The log-scale IV curve shows the leakage current caused by tail states, indicating how these states contribute to increased carrier leakage under forward bias.}
\label{figure_7}
\end{figure}

\begin{figure}[tb]
\includegraphics[width=7cm]{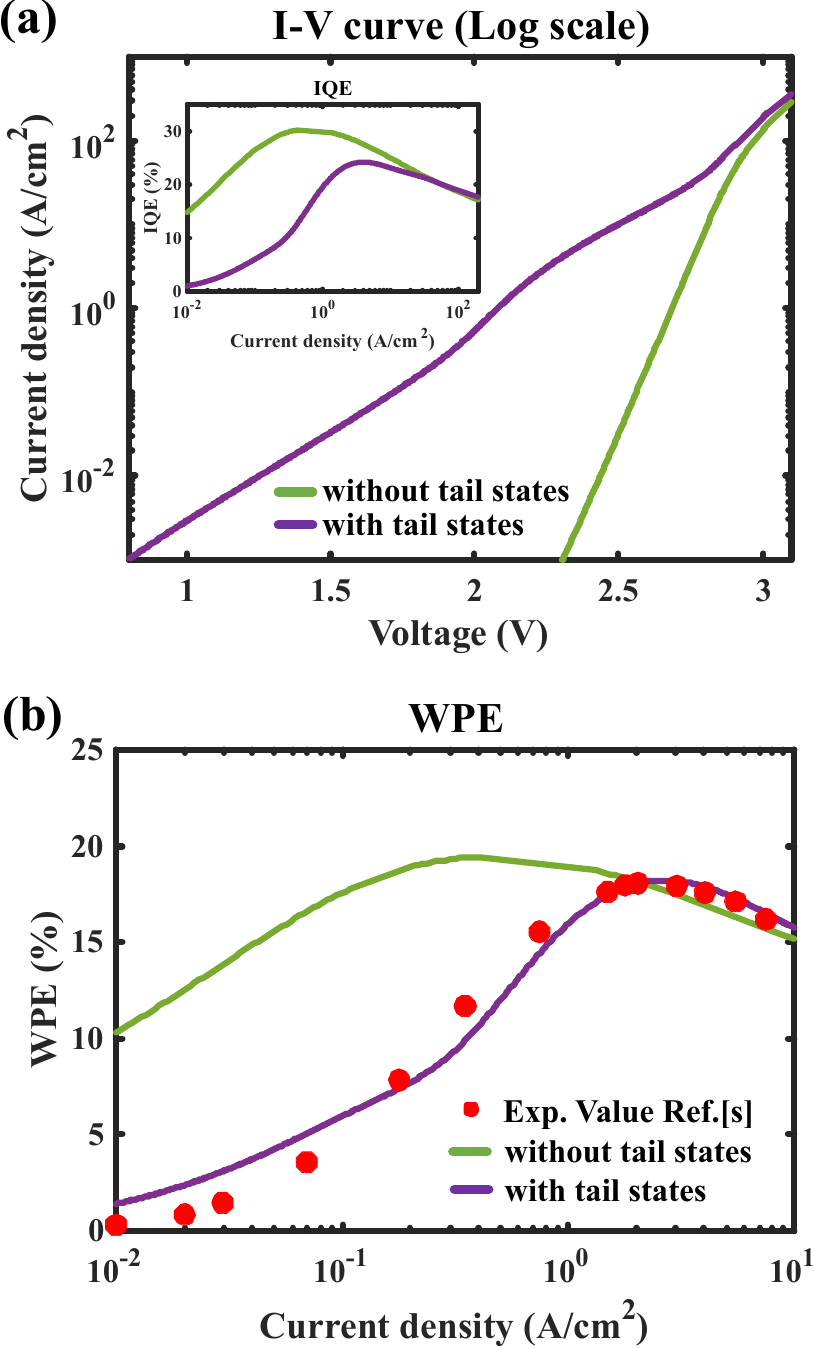}
\caption{(a) Expanded Log(I)-V plot of with/without the tail states leakage model. The tail states model can explain the leakage behavior of the LED structure. It also enhances current injection into QWs at V < 2.6V. (b)  Plot of WPE as a function of current with or without tail states. Tail states decrease WPE at low voltage as injection occurs in the TD region, with trapping by nonradiative recombination centers at dislocations. At increased voltage, injection occurs in the QW regions, thus almost recovering the original WPE without tail states. Overall, the peak WPE penalty is quite limited. The experimental value (Ref.[s]) is from the Ref.[ \cite{experiment}].}
\label{figure_8}
\end{figure}

When the voltage is lowered to 2.66V, fewer carriers are injected in the third model due to the absence of defect-mediated pathways, requiring a higher voltage to achieve comparable injection levels (Fig. \ref{figure_5}(b, e)). However, in the fourth model, which includes tail states, the current density reaches 20 A/cm\(^2\) at this voltage (Fig. \ref{figure_5}(c, f)). These tail states are incorporated into the V-defect region in the fourth model, with \(N_{tail}\) being higher at the defect center (the location of the dislocation line, Fig. \ref{figure_3}(b)). The higher current densities in the TD area (Figs. \ref{figure_5}(c) and \ref{figure_5}(f)) indicate that these tail states increase the current, particularly at lower voltages, while also leading to greater nonradiative recombination, as the TD area contains fast nonradiative recombination traps. The tail states also create leakage paths toward nonradiative centers, reducing overall efficiency. However, they facilitate carrier injection into the QWs at low voltages, as some leakage current diffuses into the QWs. Once carriers begin injecting into the c-plane quantum well (QW), they screen the polarization field, particularly at the corners of the c-plane/semipolar plane interface. This screening effect reduces the injection barrier in the c-plane region. Comparing Figs. \ref{figure_5}(e) and \ref{figure_5}(f) for hole injection at the same voltage, we observe slightly stronger vertical injection in the c-plane or semipolar plane, especially at the QW corners. Additionally, the J-V curves in Figs. \ref{figure_7} and \ref{figure_8} indicate that this change is not limited to leakage current below the turn-on voltage: the entire J-V curve shifts to a lower voltage.

To further analyze these effects, the current density-voltage (J-V) characteristics were evaluated for the four models, as shown in Fig. \ref{figure_7}(a). The forward voltage \(V_{for}\) decreases as more mechanisms are incorporated, indicating improved carrier injection efficiency. Random alloy fluctuations alone lead to a noticeable reduction in \(V_{for}\) by smoothing potential barriers, while the addition of V-defects further enhances injection efficiency by introducing additional low-energy pathways through the sidewall QWs. Tail states, however, have a more complex impact, as shown in the log-scale IV curve (Fig. \ref{figure_7}(b)). While they assist carrier injection at lower voltages by providing leakage paths, they also increase nonradiative recombination, particularly in the TD regions. Overall, the V-defect and additional tail state models help reduce the operating voltage but decrease the IQE when these defects are present. The influence of tail states can be further examined in the log(J)-V curve, as shown in Fig. \ref{figure_8}(a). Many experimental studies \cite{leakage2, leakage1, leakage3, leakage4, leakage5} report significant leakage current for voltages smaller than \(\frac{\hbar\omega}{q}\). For example, Ref.~[\cite{leakage6}] reports a sub turn-on voltage slope of approximately 393 mV/dec below the turn-on voltage. This low-voltage current behavior can be well represented by implementing the tail state model, as shown in Fig. \ref{figure_8}(a), by choosing appropriate values for \(N_{tail}\) and \(\sigma_{tail}\) in Eqn. \eqref{eq12}. While growing epitaxial layers with more defects can reduce the operating voltage, it also increases nonradiative losses. Hence, a balance must be struck between IQE and voltage to optimize WPE.

The WPE curve in Fig. \ref{figure_8}(b), assuming 85\% light extraction efficiency, shows that at low voltage, WPE is significantly reduced as most carrier injection occurs in the region with tail states, i.e., the TD region. However, at higher voltages, due to the screening effect in the QWs caused by leakage carriers, injection shifts to regions farther from the TD, almost recovering the WPE value observed without tail states.

\begin{figure}[tb]
\includegraphics[width=\columnwidth]{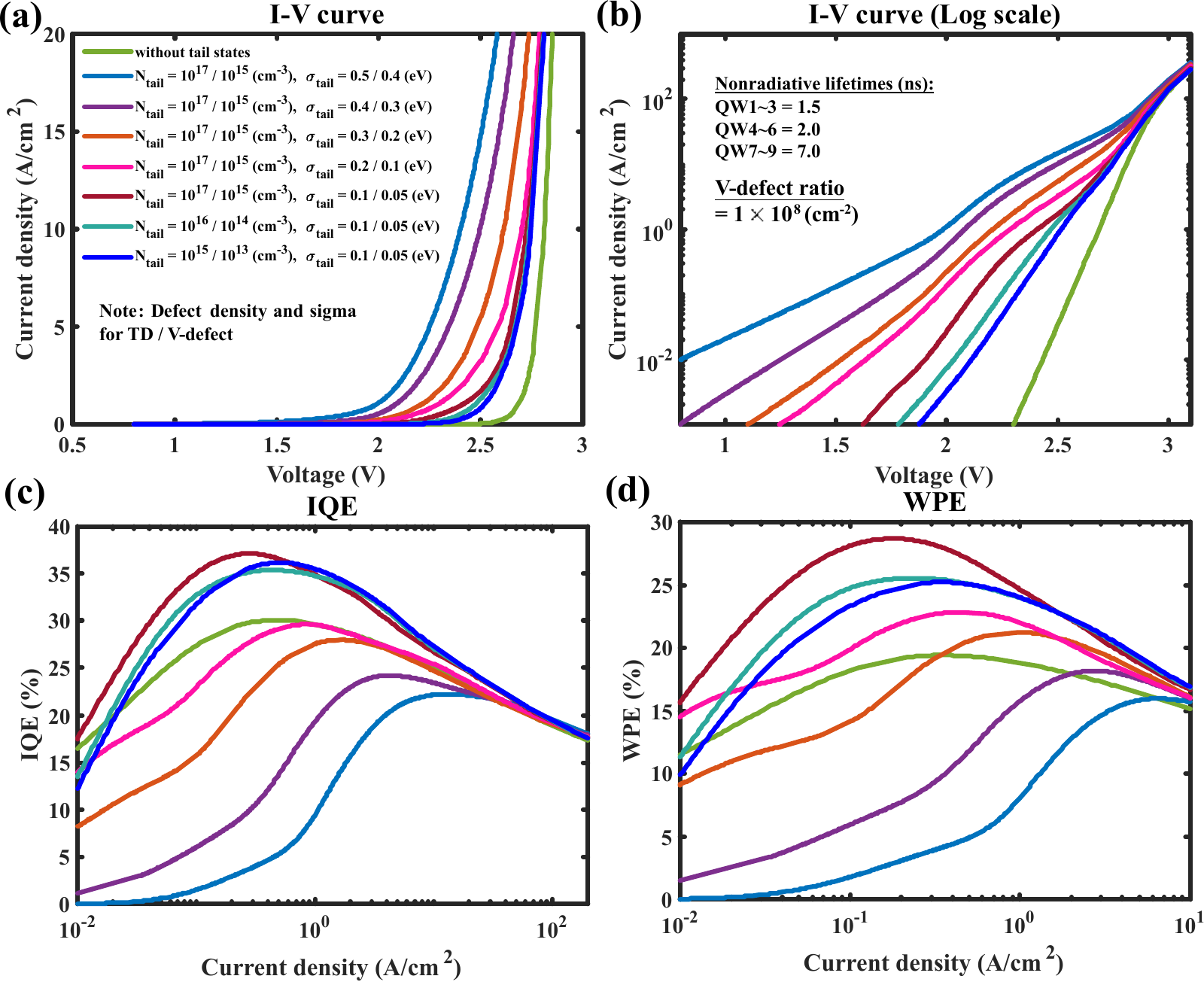}
\caption{(a) and (b) are the linear and log scale IV curves of different tail state Gaussian width $\sigma_{tail}$ and $N_{tail}$ in QB region in TD center or rest V-defect region as defined Fig.  \ref{figure_3}(b). The tail state setting in QW is kept the same. The reference cases have no tail state in either the QW or QB region. These curves show that increasing $\sigma_{tail}$ or $N_{tail}$ results in a lower turn-on voltage. (c) and (d) are the IQE and WPE curves versus current density, respectively. The larger $\sigma_{tail}$ or $N_{tail}$ leads to a smaller IQE. However, a smaller voltage helps slightly with WPE.}
\label{figure_9}
\end{figure}

\begin{figure}[htb]
\includegraphics[width=\columnwidth]{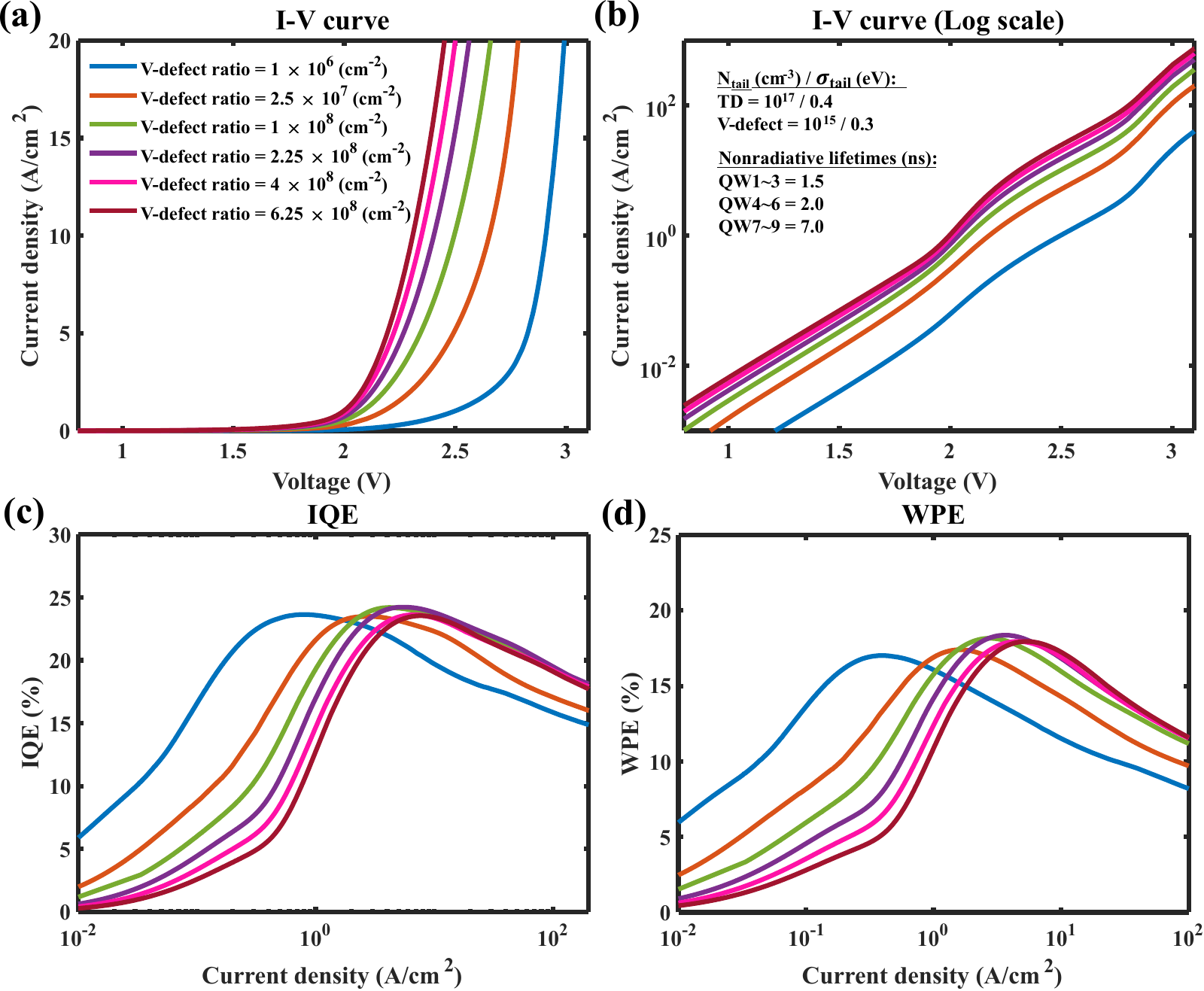}
\caption{(a) and (b) I–V curves in linear and log scale, respectively, for varying V-defect densities, showing a lower forward voltage at higher defect densities. (c) and (d) are the dependence of IQE and WPE on current density, respectively. Moderate V-defect density enhances efficiency, whereas excessive density leads to nonradiative losses.}
\label{figure_10}
\end{figure}

\begin{figure}[tb]
\includegraphics[width=\columnwidth]{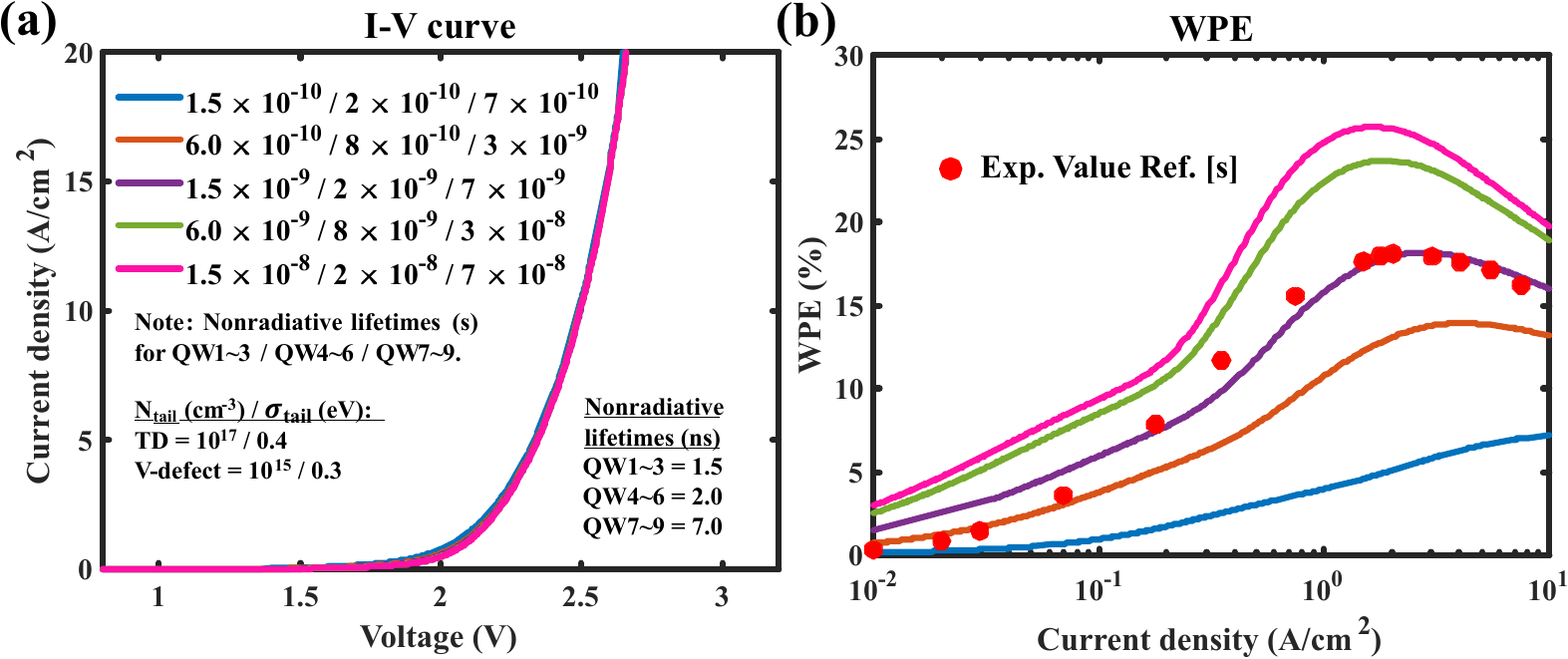}
\caption{ (a) IV curves for different nonradiative lifetime, showing that the forward voltage remains nearly unchanged, indicating minimal impact on carrier injection. The three values on nonradiative lifetime are for QW1-3, QW 4-6, and QW 7-9, respectively. (b) WPE as a function of current density, demonstrates greater sensitivity to lifetime variations. The experimental value (Ref.[s]) is from the Ref.[\cite{experiment}].}
\label{figure_11}
\end{figure}

As discussed in Fig. \ref{figure_8}, suitable parameters are adjusted to fit the experimental result closely, especially in the low current density region with a reasonable voltage and IQE. Discussing the influence of these key parameters on the result is important to understanding physics further. As shown in Fig. \ref{figure_9}(a), we change the broadening term of \(\sigma_{tail}\) in QB in analyzing the influence of leakage current. The tail state setting in QW is kept the same for comparison. As shown in Fig. \ref{figure_9}, the reduction of \(\sigma_{tail}\) in the TD and V-defect region affects the distribution of tail states, which makes the leakage smaller in the low current density region. The reduction of \(N_{tail}\) gives similar effects. Hence, to fit different results from different growth conditions, we will need to adjust these parameters slightly. Since the TD center is the major nonradiative recombination center, adding a large \(\sigma_{tail}\) will provide a path for current injection and lead to a larger nonradiative recombination and smaller IQE as shown in Fig. \ref{figure_9}(c). The IQE becomes higher as \(\sigma_{tail}\) decreases because of less carrier injection into the TD center. However, if we calculate the WPE related to the voltage, the case without tail states has a smaller WPE due to a larger forward voltage. When \(\sigma_{tail}\) in QB region is small, further reducing $N_{tail}$ continues to increase the forward voltage. The influence on IQE is slightly changed since the ratio of carrier injection into different QWs changes.

We further investigate the influence of V-defect density, including the tail states as shown in Fig. \ref{figure_10}. We can find that smaller defect density has a larger voltage due to less leakage path from the sidewall of V-defect and tail states. Because the current is crowded near the V-defect, we can see an earlier droop behavior due to the Auger recombination from the current crowding. More V-defects help the current injection and diminish current crowding. The forward voltage is also smaller. However, if the V-defect density is too high, the active c-plane area will become smaller, and there will be more nonradiative recombination centers, which will decrease the IQE again. Hence, the WPE reaches the maximum at V-defect density around 1-2 $\times 10^{8}$ $\mathrm{cm}^{-2}$. Similar observation has been reported in the early work for blue and green LEDs \cite{blueLED2}

Finally, the nonradiative lifetime resulting from the material growth condition plays a key role in WPE or IQE, as shown in Fig. \ref{figure_11}. A shorter nonradiative lifetime will lead to a smaller IQE but does not affect the I-V curve significantly. The nonradiative lifetime may result from point defect densities. The different designs of super lattices for the pre-strained layer to open the V-defect \cite{JIANG2019120} may also lead to different material quality, which all present different lifetime. These will affect the IQE of the device directly.

\section{Conclusion}
In this study, the effects of V-defects, random alloy fluctuations, and tail states were simultaneously considered to understand carrier injection and the low turn-on voltage performance of red InGaN LEDs. The inclusion of leakage currents and nonradiative recombination losses greatly improved the agreement between simulation results and experimental data compared to models considering only alloy disorder and V-defects. The tail state model not only explained the low forward voltage but also provided insight into enhanced nonradiative recombination in low-current-density regions. However, the penalty in peak WPE was found to be minimal. The random alloy, V-defect, and tail state models together help elucidate the physics of LED devices and may guide efforts to improve red InGaN LED performance by optimizing the balance between defect densities and carrier transport mechanisms to maximize both IQE and turn-on voltage, and thus WPE. These findings should provide a valuable foundation for future advancements in high-efficiency LED applications and designs.

\section{SUPPLEMENTARY MATERIAL}
Supplementary material associated with this article includes: (1) sensitivity analyses of simulated IQE to non-radiative lifetimes (Figs. S1-S4); (2) methodology for indium composition determination (Fig. S5) and its impact on emission (Fig. S6); and (3) a detailed simulation execution guide with input files. All supplementary files are available online at [URL will be inserted by publisher]. 

\begin{acknowledgments}
This project is supported by National Science and Technology Council (NSTC) under grant No. 112-2221-E-002-215-MY3, 112-2221-E-002-214-MY3, 113-2640-E-002-005, and 113-2124-M-002-013-MY3 and by Ministry of Education UAAT-KOOU project (113M7056). Support at UCSB was provided by the Solid State Lighting and Energy Electronics Center (SSLEEC); U.S. Department of Energy under the Office of Energy Efficiency \& Renewable Energy (EERE) Award No. DE-EE0009691; the National Science Foundation (NSF) RAISE program (Grant No. DMS-1839077); the Simons Foundation (Grant \#s 601952 and 1027114 for JSS and CW, respectively). 
\end{acknowledgments}

\section*{AUTHOR DECLARATIONS}
Huai-Chin Huang: conceptualization (equal); data curation (lead); formal analysis (equal); methodology (support); validation (lead); visualization(lead); writing – original draft (lead). 

Shih-Min Chen: conceptualization (support); data curation (support); formal analysis (equal); validation (support). 

Claude Weisbuch: conceptualization (equal);writing-review and editing (equal).  supervision (support)

James S. Speck: conceptualization (equal);writing-review and editing (equal). funding acquisition (equal);  supervision (support);

Yuh-Renn Wu: conceptualization (equal); formal analysis (equal); funding acquisition (lead); project administration (lead); software(lead); supervision (lead); writing-review and editing (lead).
\section*{Conflict of interest}
The authors have no conflicts to disclose

\section*{Data Availability Statement}
The data that support the findings of this study are available from the corresponding author upon reasonable request.

\section{References}
\bibliography{Red_LED_reference}

\begin{thebibliography}{56}%
\makeatletter
\providecommand \@ifxundefined [1]{%
 \@ifx{#1\undefined}
}%
\providecommand \@ifnum [1]{%
 \ifnum #1\expandafter \@firstoftwo
 \else \expandafter \@secondoftwo
 \fi
}%
\providecommand \@ifx [1]{%
 \ifx #1\expandafter \@firstoftwo
 \else \expandafter \@secondoftwo
 \fi
}%
\providecommand \natexlab [1]{#1}%
\providecommand \enquote  [1]{``#1''}%
\providecommand \bibnamefont  [1]{#1}%
\providecommand \bibfnamefont [1]{#1}%
\providecommand \citenamefont [1]{#1}%
\providecommand \href@noop [0]{\@secondoftwo}%
\providecommand \href [0]{\begingroup \@sanitize@url \@href}%
\providecommand \@href[1]{\@@startlink{#1}\@@href}%
\providecommand \@@href[1]{\endgroup#1\@@endlink}%
\providecommand \@sanitize@url [0]{\catcode `\\12\catcode `\$12\catcode
  `\&12\catcode `\#12\catcode `\^12\catcode `\_12\catcode `\%12\relax}%
\providecommand \@@startlink[1]{}%
\providecommand \@@endlink[0]{}%
\providecommand \url  [0]{\begingroup\@sanitize@url \@url }%
\providecommand \@url [1]{\endgroup\@href {#1}{\urlprefix }}%
\providecommand \urlprefix  [0]{URL }%
\providecommand \Eprint [0]{\href }%
\providecommand \doibase [0]{http://dx.doi.org/}%
\providecommand \selectlanguage [0]{\@gobble}%
\providecommand \bibinfo  [0]{\@secondoftwo}%
\providecommand \bibfield  [0]{\@secondoftwo}%
\providecommand \translation [1]{[#1]}%
\providecommand \BibitemOpen [0]{}%
\providecommand \bibitemStop [0]{}%
\providecommand \bibitemNoStop [0]{.\EOS\space}%
\providecommand \EOS [0]{\spacefactor3000\relax}%
\providecommand \BibitemShut  [1]{\csname bibitem#1\endcsname}%
\let\auto@bib@innerbib\@empty
\bibitem [{\citenamefont {Day}\ \emph {et~al.}(2011)\citenamefont {Day},
  \citenamefont {Li}, \citenamefont {Lie}, \citenamefont {Bradford},
  \citenamefont {Lin},\ and\ \citenamefont {Jiang}}]{microLED3}%
  \BibitemOpen
  \bibfield  {author} {\bibinfo {author} {\bibfnamefont {J.}~\bibnamefont
  {Day}}, \bibinfo {author} {\bibfnamefont {J.}~\bibnamefont {Li}}, \bibinfo
  {author} {\bibfnamefont {D.}~\bibnamefont {Lie}}, \bibinfo {author}
  {\bibfnamefont {C.}~\bibnamefont {Bradford}}, \bibinfo {author}
  {\bibfnamefont {J.}~\bibnamefont {Lin}}, \ and\ \bibinfo {author}
  {\bibfnamefont {H.}~\bibnamefont {Jiang}},\ }\bibfield  {title} {\enquote
  {\bibinfo {title} {{III-Nitride} full-scale high-resolution microdisplays},}\
  }\href@noop {} {\bibfield  {journal} {\bibinfo  {journal} {Applied Physics
  Letters}\ }\textbf {\bibinfo {volume} {99}},\ \bibinfo {pages} {031116}
  (\bibinfo {year} {2011})}\BibitemShut {NoStop}%
\bibitem [{\citenamefont {Lin}\ and\ \citenamefont {Jiang}(2020)}]{microLED4}%
  \BibitemOpen
  \bibfield  {author} {\bibinfo {author} {\bibfnamefont {J.}~\bibnamefont
  {Lin}}\ and\ \bibinfo {author} {\bibfnamefont {H.}~\bibnamefont {Jiang}},\
  }\bibfield  {title} {\enquote {\bibinfo {title} {Development of
  {microLED}},}\ }\href@noop {} {\bibfield  {journal} {\bibinfo  {journal}
  {Applied Physics Letters}\ }\textbf {\bibinfo {volume} {116}},\ \bibinfo
  {pages} {100502} (\bibinfo {year} {2020})}\BibitemShut {NoStop}%
\bibitem [{\citenamefont {Pandey}\ and\ \citenamefont {Mi}(2022)}]{microLED5}%
  \BibitemOpen
  \bibfield  {author} {\bibinfo {author} {\bibfnamefont {A.}~\bibnamefont
  {Pandey}}\ and\ \bibinfo {author} {\bibfnamefont {Z.}~\bibnamefont {Mi}},\
  }\bibfield  {title} {\enquote {\bibinfo {title} {{III-nitride} nanostructures
  for high efficiency {micro-LEDs} and ultraviolet optoelectronics},}\
  }\href@noop {} {\bibfield  {journal} {\bibinfo  {journal} {IEEE Journal of
  Quantum Electronics}\ }\textbf {\bibinfo {volume} {58}},\ \bibinfo {pages}
  {1--13} (\bibinfo {year} {2022})}\BibitemShut {NoStop}%
\bibitem [{\citenamefont {Lin}\ \emph {et~al.}(2023)\citenamefont {Lin},
  \citenamefont {Wu}, \citenamefont {Kuo}, \citenamefont {Wong}, \citenamefont
  {DenBaars}, \citenamefont {Nakamura}, \citenamefont {Pandey}, \citenamefont
  {Mi}, \citenamefont {Tian}, \citenamefont {Ohkawa} \emph
  {et~al.}}]{microLED2}%
  \BibitemOpen
  \bibfield  {author} {\bibinfo {author} {\bibfnamefont {C.-C.}\ \bibnamefont
  {Lin}}, \bibinfo {author} {\bibfnamefont {Y.-R.}\ \bibnamefont {Wu}},
  \bibinfo {author} {\bibfnamefont {H.-C.}\ \bibnamefont {Kuo}}, \bibinfo
  {author} {\bibfnamefont {M.~S.}\ \bibnamefont {Wong}}, \bibinfo {author}
  {\bibfnamefont {S.~P.}\ \bibnamefont {DenBaars}}, \bibinfo {author}
  {\bibfnamefont {S.}~\bibnamefont {Nakamura}}, \bibinfo {author}
  {\bibfnamefont {A.}~\bibnamefont {Pandey}}, \bibinfo {author} {\bibfnamefont
  {Z.}~\bibnamefont {Mi}}, \bibinfo {author} {\bibfnamefont {P.}~\bibnamefont
  {Tian}}, \bibinfo {author} {\bibfnamefont {K.}~\bibnamefont {Ohkawa}},  \emph
  {et~al.},\ }\bibfield  {title} {\enquote {\bibinfo {title} {The {micro-LED}
  roadmap: status quo and prospects},}\ }\href@noop {} {\bibfield  {journal}
  {\bibinfo  {journal} {Journal of Physics: Photonics}\ }\textbf {\bibinfo
  {volume} {5}},\ \bibinfo {pages} {042502} (\bibinfo {year}
  {2023})}\BibitemShut {NoStop}%
\bibitem [{\citenamefont {Weisbuch}(2019)}]{blueLED}%
  \BibitemOpen
  \bibfield  {author} {\bibinfo {author} {\bibfnamefont {C.}~\bibnamefont
  {Weisbuch}},\ }\bibfield  {title} {\enquote {\bibinfo {title} {On the search
  for efficient solid state light emitters: {Past}, present, future},}\
  }\href@noop {} {\bibfield  {journal} {\bibinfo  {journal} {ECS Journal of
  Solid State Science and Technology}\ }\textbf {\bibinfo {volume} {9}},\
  \bibinfo {pages} {016022} (\bibinfo {year} {2019})}\BibitemShut {NoStop}%
\bibitem [{\citenamefont {Ho}\ \emph {et~al.}(2022)\citenamefont {Ho},
  \citenamefont {Speck}, \citenamefont {Weisbuch},\ and\ \citenamefont
  {Wu}}]{blueLED2}%
  \BibitemOpen
  \bibfield  {author} {\bibinfo {author} {\bibfnamefont {C.-H.}\ \bibnamefont
  {Ho}}, \bibinfo {author} {\bibfnamefont {J.~S.}\ \bibnamefont {Speck}},
  \bibinfo {author} {\bibfnamefont {C.}~\bibnamefont {Weisbuch}}, \ and\
  \bibinfo {author} {\bibfnamefont {Y.-R.}\ \bibnamefont {Wu}},\ }\bibfield
  {title} {\enquote {\bibinfo {title} {Efficiency and forward voltage of blue
  and green lateral {LEDs} with {V-shaped} defects and random alloy fluctuation
  in quantum wells},}\ }\href@noop {} {\bibfield  {journal} {\bibinfo
  {journal} {Physical Review Applied}\ }\textbf {\bibinfo {volume} {17}},\
  \bibinfo {pages} {014033} (\bibinfo {year} {2022})}\BibitemShut {NoStop}%
\bibitem [{\citenamefont {Massabuau}\ \emph {et~al.}(2014)\citenamefont
  {Massabuau}, \citenamefont {Davies}, \citenamefont {Oehler}, \citenamefont
  {Pamenter}, \citenamefont {Thrush}, \citenamefont {Kappers}, \citenamefont
  {Kovacs}, \citenamefont {Williams}, \citenamefont {Hopkins}, \citenamefont
  {Humphreys} \emph {et~al.}}]{greengap}%
  \BibitemOpen
  \bibfield  {author} {\bibinfo {author} {\bibfnamefont {F.-P.}\ \bibnamefont
  {Massabuau}}, \bibinfo {author} {\bibfnamefont {M.~J.}\ \bibnamefont
  {Davies}}, \bibinfo {author} {\bibfnamefont {F.}~\bibnamefont {Oehler}},
  \bibinfo {author} {\bibfnamefont {S.}~\bibnamefont {Pamenter}}, \bibinfo
  {author} {\bibfnamefont {E.}~\bibnamefont {Thrush}}, \bibinfo {author}
  {\bibfnamefont {M.~J.}\ \bibnamefont {Kappers}}, \bibinfo {author}
  {\bibfnamefont {A.}~\bibnamefont {Kovacs}}, \bibinfo {author} {\bibfnamefont
  {T.}~\bibnamefont {Williams}}, \bibinfo {author} {\bibfnamefont {M.~A.}\
  \bibnamefont {Hopkins}}, \bibinfo {author} {\bibfnamefont {C.~J.}\
  \bibnamefont {Humphreys}},  \emph {et~al.},\ }\bibfield  {title} {\enquote
  {\bibinfo {title} {The impact of trench defects in {InGaN/GaN} light emitting
  diodes and implications for the “green gap” problem},}\ }\href@noop {}
  {\bibfield  {journal} {\bibinfo  {journal} {Applied Physics Letters}\
  }\textbf {\bibinfo {volume} {105}},\ \bibinfo {pages} {112110} (\bibinfo
  {year} {2014})}\BibitemShut {NoStop}%
\bibitem [{\citenamefont {Zhang}\ \emph {et~al.}(2020)\citenamefont {Zhang},
  \citenamefont {Zhang}, \citenamefont {Gao}, \citenamefont {Wang},
  \citenamefont {Zheng}, \citenamefont {Zhang}, \citenamefont {Wu},
  \citenamefont {Xu}, \citenamefont {Ding}, \citenamefont {Quan} \emph
  {et~al.}}]{experiment}%
  \BibitemOpen
  \bibfield  {author} {\bibinfo {author} {\bibfnamefont {S.}~\bibnamefont
  {Zhang}}, \bibinfo {author} {\bibfnamefont {J.}~\bibnamefont {Zhang}},
  \bibinfo {author} {\bibfnamefont {J.}~\bibnamefont {Gao}}, \bibinfo {author}
  {\bibfnamefont {X.}~\bibnamefont {Wang}}, \bibinfo {author} {\bibfnamefont
  {C.}~\bibnamefont {Zheng}}, \bibinfo {author} {\bibfnamefont
  {M.}~\bibnamefont {Zhang}}, \bibinfo {author} {\bibfnamefont
  {X.}~\bibnamefont {Wu}}, \bibinfo {author} {\bibfnamefont {L.}~\bibnamefont
  {Xu}}, \bibinfo {author} {\bibfnamefont {J.}~\bibnamefont {Ding}}, \bibinfo
  {author} {\bibfnamefont {Z.}~\bibnamefont {Quan}},  \emph {et~al.},\
  }\bibfield  {title} {\enquote {\bibinfo {title} {Efficient emission of
  {InGaN}-based light-emitting diodes: toward orange and red},}\ }\href@noop {}
  {\bibfield  {journal} {\bibinfo  {journal} {Photonics Research}\ }\textbf
  {\bibinfo {volume} {8}},\ \bibinfo {pages} {1671--1675} (\bibinfo {year}
  {2020})}\BibitemShut {NoStop}%
\bibitem [{\citenamefont {Nath}\ \emph {et~al.}(2013)\citenamefont {Nath},
  \citenamefont {Yang}, \citenamefont {Lee}, \citenamefont {Park},
  \citenamefont {Wu},\ and\ \citenamefont {Rajan}}]{polarization4}%
  \BibitemOpen
  \bibfield  {author} {\bibinfo {author} {\bibfnamefont {D.}~\bibnamefont
  {Nath}}, \bibinfo {author} {\bibfnamefont {Z.}~\bibnamefont {Yang}}, \bibinfo
  {author} {\bibfnamefont {C.-Y.}\ \bibnamefont {Lee}}, \bibinfo {author}
  {\bibfnamefont {P.}~\bibnamefont {Park}}, \bibinfo {author} {\bibfnamefont
  {Y.-R.}\ \bibnamefont {Wu}}, \ and\ \bibinfo {author} {\bibfnamefont
  {S.}~\bibnamefont {Rajan}},\ }\bibfield  {title} {\enquote {\bibinfo {title}
  {Unipolar vertical transport in {GaN/AlGaN/GaN} heterostructures},}\
  }\href@noop {} {\bibfield  {journal} {\bibinfo  {journal} {Applied Physics
  Letters}\ }\textbf {\bibinfo {volume} {103}},\ \bibinfo {pages} {022102}
  (\bibinfo {year} {2013})}\BibitemShut {NoStop}%
\bibitem [{\citenamefont {Lheureux}\ \emph {et~al.}(2020)\citenamefont
  {Lheureux}, \citenamefont {Lynsky}, \citenamefont {Wu}, \citenamefont
  {Speck},\ and\ \citenamefont {Weisbuch}}]{polarization1}%
  \BibitemOpen
  \bibfield  {author} {\bibinfo {author} {\bibfnamefont {G.}~\bibnamefont
  {Lheureux}}, \bibinfo {author} {\bibfnamefont {C.}~\bibnamefont {Lynsky}},
  \bibinfo {author} {\bibfnamefont {Y.-R.}\ \bibnamefont {Wu}}, \bibinfo
  {author} {\bibfnamefont {J.~S.}\ \bibnamefont {Speck}}, \ and\ \bibinfo
  {author} {\bibfnamefont {C.}~\bibnamefont {Weisbuch}},\ }\bibfield  {title}
  {\enquote {\bibinfo {title} {A {3D} simulation comparison of carrier
  transport in green and blue c-plane multi-quantum well nitride light emitting
  diodes},}\ }\href@noop {} {\bibfield  {journal} {\bibinfo  {journal} {Journal
  of Applied Physics}\ }\textbf {\bibinfo {volume} {128}},\ \bibinfo {pages}
  {235703} (\bibinfo {year} {2020})}\BibitemShut {NoStop}%
\bibitem [{\citenamefont {Lynsky}\ \emph {et~al.}(2020)\citenamefont {Lynsky},
  \citenamefont {Alhassan}, \citenamefont {Lheureux}, \citenamefont {Bonef},
  \citenamefont {DenBaars}, \citenamefont {Nakamura}, \citenamefont {Wu},
  \citenamefont {Weisbuch},\ and\ \citenamefont {Speck}}]{polarization2}%
  \BibitemOpen
  \bibfield  {author} {\bibinfo {author} {\bibfnamefont {C.}~\bibnamefont
  {Lynsky}}, \bibinfo {author} {\bibfnamefont {A.~I.}\ \bibnamefont
  {Alhassan}}, \bibinfo {author} {\bibfnamefont {G.}~\bibnamefont {Lheureux}},
  \bibinfo {author} {\bibfnamefont {B.}~\bibnamefont {Bonef}}, \bibinfo
  {author} {\bibfnamefont {S.~P.}\ \bibnamefont {DenBaars}}, \bibinfo {author}
  {\bibfnamefont {S.}~\bibnamefont {Nakamura}}, \bibinfo {author}
  {\bibfnamefont {Y.-R.}\ \bibnamefont {Wu}}, \bibinfo {author} {\bibfnamefont
  {C.}~\bibnamefont {Weisbuch}}, \ and\ \bibinfo {author} {\bibfnamefont
  {J.~S.}\ \bibnamefont {Speck}},\ }\bibfield  {title} {\enquote {\bibinfo
  {title} {Barriers to carrier transport in multiple quantum well nitride-based
  c-plane green light emitting diodes},}\ }\href@noop {} {\bibfield  {journal}
  {\bibinfo  {journal} {Physical Review Materials}\ }\textbf {\bibinfo {volume}
  {4}},\ \bibinfo {pages} {054604} (\bibinfo {year} {2020})}\BibitemShut
  {NoStop}%
\bibitem [{\citenamefont {Qwah}\ \emph {et~al.}(2020)\citenamefont {Qwah},
  \citenamefont {Monavarian}, \citenamefont {Lheureux}, \citenamefont {Wang},
  \citenamefont {Wu},\ and\ \citenamefont {Speck}}]{polarization3}%
  \BibitemOpen
  \bibfield  {author} {\bibinfo {author} {\bibfnamefont {K.}~\bibnamefont
  {Qwah}}, \bibinfo {author} {\bibfnamefont {M.}~\bibnamefont {Monavarian}},
  \bibinfo {author} {\bibfnamefont {G.}~\bibnamefont {Lheureux}}, \bibinfo
  {author} {\bibfnamefont {J.}~\bibnamefont {Wang}}, \bibinfo {author}
  {\bibfnamefont {Y.-R.}\ \bibnamefont {Wu}}, \ and\ \bibinfo {author}
  {\bibfnamefont {J.}~\bibnamefont {Speck}},\ }\bibfield  {title} {\enquote
  {\bibinfo {title} {Theoretical and experimental investigations of vertical
  hole transport through unipolar {AlGaN} structures: Impacts of random alloy
  disorder},}\ }\href@noop {} {\bibfield  {journal} {\bibinfo  {journal}
  {Applied Physics Letters}\ }\textbf {\bibinfo {volume} {117}},\ \bibinfo
  {pages} {022107} (\bibinfo {year} {2020})}\BibitemShut {NoStop}%
\bibitem [{\citenamefont {Li}\ \emph {et~al.}(2017)\citenamefont {Li},
  \citenamefont {Piccardo}, \citenamefont {Lu}, \citenamefont {Mayboroda},
  \citenamefont {Martinelli}, \citenamefont {Peretti}, \citenamefont {Speck},
  \citenamefont {Weisbuch}, \citenamefont {Filoche},\ and\ \citenamefont
  {Wu}}]{LL2}%
  \BibitemOpen
  \bibfield  {author} {\bibinfo {author} {\bibfnamefont {C.-K.}\ \bibnamefont
  {Li}}, \bibinfo {author} {\bibfnamefont {M.}~\bibnamefont {Piccardo}},
  \bibinfo {author} {\bibfnamefont {L.-S.}\ \bibnamefont {Lu}}, \bibinfo
  {author} {\bibfnamefont {S.}~\bibnamefont {Mayboroda}}, \bibinfo {author}
  {\bibfnamefont {L.}~\bibnamefont {Martinelli}}, \bibinfo {author}
  {\bibfnamefont {J.}~\bibnamefont {Peretti}}, \bibinfo {author} {\bibfnamefont
  {J.~S.}\ \bibnamefont {Speck}}, \bibinfo {author} {\bibfnamefont
  {C.}~\bibnamefont {Weisbuch}}, \bibinfo {author} {\bibfnamefont
  {M.}~\bibnamefont {Filoche}}, \ and\ \bibinfo {author} {\bibfnamefont
  {Y.-R.}\ \bibnamefont {Wu}},\ }\bibfield  {title} {\enquote {\bibinfo {title}
  {Localization landscape theory of disorder in semiconductors. {III.}
  {Application} to carrier transport and recombination in light emitting
  diodes},}\ }\href@noop {} {\bibfield  {journal} {\bibinfo  {journal}
  {Physical Review B}\ }\textbf {\bibinfo {volume} {95}},\ \bibinfo {pages}
  {144206} (\bibinfo {year} {2017})}\BibitemShut {NoStop}%
\bibitem [{\citenamefont {Wu}\ \emph {et~al.}(1998)\citenamefont {Wu},
  \citenamefont {Elsass}, \citenamefont {Abare}, \citenamefont {Mack},
  \citenamefont {Keller}, \citenamefont {Petroff}, \citenamefont {DenBaars},
  \citenamefont {Speck},\ and\ \citenamefont {Rosner}}]{Vdefect1}%
  \BibitemOpen
  \bibfield  {author} {\bibinfo {author} {\bibfnamefont {X.}~\bibnamefont
  {Wu}}, \bibinfo {author} {\bibfnamefont {C.}~\bibnamefont {Elsass}}, \bibinfo
  {author} {\bibfnamefont {A.}~\bibnamefont {Abare}}, \bibinfo {author}
  {\bibfnamefont {M.}~\bibnamefont {Mack}}, \bibinfo {author} {\bibfnamefont
  {S.}~\bibnamefont {Keller}}, \bibinfo {author} {\bibfnamefont
  {P.}~\bibnamefont {Petroff}}, \bibinfo {author} {\bibfnamefont
  {S.}~\bibnamefont {DenBaars}}, \bibinfo {author} {\bibfnamefont
  {J.}~\bibnamefont {Speck}}, \ and\ \bibinfo {author} {\bibfnamefont
  {S.}~\bibnamefont {Rosner}},\ }\bibfield  {title} {\enquote {\bibinfo {title}
  {Structural origin of {V-defects} and correlation with localized excitonic
  centers in {InGaN/GaN} multiple quantum wells},}\ }\href@noop {} {\bibfield
  {journal} {\bibinfo  {journal} {Applied Physics Letters}\ }\textbf {\bibinfo
  {volume} {72}},\ \bibinfo {pages} {692--694} (\bibinfo {year}
  {1998})}\BibitemShut {NoStop}%
\bibitem [{\citenamefont {Yang}\ \emph {et~al.}(2014)\citenamefont {Yang},
  \citenamefont {Shivaraman}, \citenamefont {Speck},\ and\ \citenamefont
  {Wu}}]{randomalloyfluctuation1}%
  \BibitemOpen
  \bibfield  {author} {\bibinfo {author} {\bibfnamefont {T.-J.}\ \bibnamefont
  {Yang}}, \bibinfo {author} {\bibfnamefont {R.}~\bibnamefont {Shivaraman}},
  \bibinfo {author} {\bibfnamefont {J.~S.}\ \bibnamefont {Speck}}, \ and\
  \bibinfo {author} {\bibfnamefont {Y.-R.}\ \bibnamefont {Wu}},\ }\bibfield
  {title} {\enquote {\bibinfo {title} {The influence of random indium alloy
  fluctuations in indium gallium nitride quantum wells on the device
  behavior},}\ }\href@noop {} {\bibfield  {journal} {\bibinfo  {journal}
  {Journal of Applied Physics}\ }\textbf {\bibinfo {volume} {116}},\ \bibinfo
  {pages} {113104} (\bibinfo {year} {2014})}\BibitemShut {NoStop}%
\bibitem [{\citenamefont {Nakamura}(1998)}]{Vdefect17}%
  \BibitemOpen
  \bibfield  {author} {\bibinfo {author} {\bibfnamefont {S.}~\bibnamefont
  {Nakamura}},\ }\bibfield  {title} {\enquote {\bibinfo {title} {The roles of
  structural imperfections in {InGaN}-based blue light-emitting diodes and
  laser diodes},}\ }\href@noop {} {\bibfield  {journal} {\bibinfo  {journal}
  {Science}\ }\textbf {\bibinfo {volume} {281}},\ \bibinfo {pages} {956--961}
  (\bibinfo {year} {1998})}\BibitemShut {NoStop}%
\bibitem [{\citenamefont {Heying}\ \emph {et~al.}(1999)\citenamefont {Heying},
  \citenamefont {Tarsa}, \citenamefont {Elsass}, \citenamefont {Fini},
  \citenamefont {DenBaars},\ and\ \citenamefont {Speck}}]{Vdefect11}%
  \BibitemOpen
  \bibfield  {author} {\bibinfo {author} {\bibfnamefont {B.}~\bibnamefont
  {Heying}}, \bibinfo {author} {\bibfnamefont {E.}~\bibnamefont {Tarsa}},
  \bibinfo {author} {\bibfnamefont {C.}~\bibnamefont {Elsass}}, \bibinfo
  {author} {\bibfnamefont {P.}~\bibnamefont {Fini}}, \bibinfo {author}
  {\bibfnamefont {S.}~\bibnamefont {DenBaars}}, \ and\ \bibinfo {author}
  {\bibfnamefont {J.}~\bibnamefont {Speck}},\ }\bibfield  {title} {\enquote
  {\bibinfo {title} {Dislocation mediated surface morphology of {GaN}},}\
  }\href@noop {} {\bibfield  {journal} {\bibinfo  {journal} {Journal of Applied
  Physics}\ }\textbf {\bibinfo {volume} {85}},\ \bibinfo {pages} {6470--6476}
  (\bibinfo {year} {1999})}\BibitemShut {NoStop}%
\bibitem [{\citenamefont {Bright}, \citenamefont {Sharma},\ and\ \citenamefont
  {Humphreys}(2001)}]{Vdefect18}%
  \BibitemOpen
  \bibfield  {author} {\bibinfo {author} {\bibfnamefont {A.~N.}\ \bibnamefont
  {Bright}}, \bibinfo {author} {\bibfnamefont {N.}~\bibnamefont {Sharma}}, \
  and\ \bibinfo {author} {\bibfnamefont {C.~J.}\ \bibnamefont {Humphreys}},\
  }\bibfield  {title} {\enquote {\bibinfo {title} {Analysis of contacts and
  {V-defects} in {GaN} device structures by transmission electron
  microscopy},}\ }\href@noop {} {\bibfield  {journal} {\bibinfo  {journal}
  {Microscopy}\ }\textbf {\bibinfo {volume} {50}},\ \bibinfo {pages} {489--495}
  (\bibinfo {year} {2001})}\BibitemShut {NoStop}%
\bibitem [{\citenamefont {Le}\ \emph {et~al.}(2013)\citenamefont {Le},
  \citenamefont {Zhao}, \citenamefont {Jiang}, \citenamefont {Li},
  \citenamefont {Wu}, \citenamefont {Chen}, \citenamefont {Liu}, \citenamefont
  {Yang}, \citenamefont {Li}, \citenamefont {He} \emph {et~al.}}]{Vdefect13}%
  \BibitemOpen
  \bibfield  {author} {\bibinfo {author} {\bibfnamefont {L.}~\bibnamefont
  {Le}}, \bibinfo {author} {\bibfnamefont {D.}~\bibnamefont {Zhao}}, \bibinfo
  {author} {\bibfnamefont {D.}~\bibnamefont {Jiang}}, \bibinfo {author}
  {\bibfnamefont {L.}~\bibnamefont {Li}}, \bibinfo {author} {\bibfnamefont
  {L.}~\bibnamefont {Wu}}, \bibinfo {author} {\bibfnamefont {P.}~\bibnamefont
  {Chen}}, \bibinfo {author} {\bibfnamefont {Z.}~\bibnamefont {Liu}}, \bibinfo
  {author} {\bibfnamefont {J.}~\bibnamefont {Yang}}, \bibinfo {author}
  {\bibfnamefont {X.}~\bibnamefont {Li}}, \bibinfo {author} {\bibfnamefont
  {X.}~\bibnamefont {He}},  \emph {et~al.},\ }\bibfield  {title} {\enquote
  {\bibinfo {title} {Effect of {V-defects} on the performance deterioration of
  {InGaN/GaN} multiple-quantum-well light-emitting diodes with varying barrier
  layer thickness},}\ }\href@noop {} {\bibfield  {journal} {\bibinfo  {journal}
  {Journal of Applied Physics}\ }\textbf {\bibinfo {volume} {114}},\ \bibinfo
  {pages} {143706} (\bibinfo {year} {2013})}\BibitemShut {NoStop}%
\bibitem [{\citenamefont {Kim}\ \emph {et~al.}(2014)\citenamefont {Kim},
  \citenamefont {Cho}, \citenamefont {Ko}, \citenamefont {Li}, \citenamefont
  {Won}, \citenamefont {Lee}, \citenamefont {Park}, \citenamefont {Kim},\ and\
  \citenamefont {Kim}}]{Vdefect16}%
  \BibitemOpen
  \bibfield  {author} {\bibinfo {author} {\bibfnamefont {J.}~\bibnamefont
  {Kim}}, \bibinfo {author} {\bibfnamefont {Y.-H.}\ \bibnamefont {Cho}},
  \bibinfo {author} {\bibfnamefont {D.-S.}\ \bibnamefont {Ko}}, \bibinfo
  {author} {\bibfnamefont {X.-S.}\ \bibnamefont {Li}}, \bibinfo {author}
  {\bibfnamefont {J.-Y.}\ \bibnamefont {Won}}, \bibinfo {author} {\bibfnamefont
  {E.}~\bibnamefont {Lee}}, \bibinfo {author} {\bibfnamefont {S.-H.}\
  \bibnamefont {Park}}, \bibinfo {author} {\bibfnamefont {J.-Y.}\ \bibnamefont
  {Kim}}, \ and\ \bibinfo {author} {\bibfnamefont {S.}~\bibnamefont {Kim}},\
  }\bibfield  {title} {\enquote {\bibinfo {title} {Influence of {V-pits} on the
  efficiency droop in {InGaN/GaN} quantum wells},}\ }\href@noop {} {\bibfield
  {journal} {\bibinfo  {journal} {Optics express}\ }\textbf {\bibinfo {volume}
  {22}},\ \bibinfo {pages} {A857--A866} (\bibinfo {year} {2014})}\BibitemShut
  {NoStop}%
\bibitem [{\citenamefont {Wang}\ \emph {et~al.}(2015)\citenamefont {Wang},
  \citenamefont {Wang}, \citenamefont {Tan},\ and\ \citenamefont
  {Zeng}}]{Vdefect12}%
  \BibitemOpen
  \bibfield  {author} {\bibinfo {author} {\bibfnamefont {H.}~\bibnamefont
  {Wang}}, \bibinfo {author} {\bibfnamefont {X.}~\bibnamefont {Wang}}, \bibinfo
  {author} {\bibfnamefont {Q.}~\bibnamefont {Tan}}, \ and\ \bibinfo {author}
  {\bibfnamefont {X.}~\bibnamefont {Zeng}},\ }\bibfield  {title} {\enquote
  {\bibinfo {title} {V-defects formation and optical properties of {InGaN/GaN}
  multiple quantum well {LED} grown on patterned sapphire substrate},}\
  }\href@noop {} {\bibfield  {journal} {\bibinfo  {journal} {Materials Science
  in Semiconductor Processing}\ }\textbf {\bibinfo {volume} {29}},\ \bibinfo
  {pages} {112--116} (\bibinfo {year} {2015})}\BibitemShut {NoStop}%
\bibitem [{\citenamefont {Nie}\ \emph {et~al.}(2018)\citenamefont {Nie},
  \citenamefont {Jiang}, \citenamefont {Gan}, \citenamefont {Liu},
  \citenamefont {Yan},\ and\ \citenamefont {Fang}}]{Vdefect19}%
  \BibitemOpen
  \bibfield  {author} {\bibinfo {author} {\bibfnamefont {Q.}~\bibnamefont
  {Nie}}, \bibinfo {author} {\bibfnamefont {Z.}~\bibnamefont {Jiang}}, \bibinfo
  {author} {\bibfnamefont {Z.}~\bibnamefont {Gan}}, \bibinfo {author}
  {\bibfnamefont {S.}~\bibnamefont {Liu}}, \bibinfo {author} {\bibfnamefont
  {H.}~\bibnamefont {Yan}}, \ and\ \bibinfo {author} {\bibfnamefont
  {H.}~\bibnamefont {Fang}},\ }\bibfield  {title} {\enquote {\bibinfo {title}
  {Defect analysis of the {LED} structure deposited on the sapphire
  substrate},}\ }\href@noop {} {\bibfield  {journal} {\bibinfo  {journal}
  {Journal of Crystal Growth}\ }\textbf {\bibinfo {volume} {488}},\ \bibinfo
  {pages} {1--7} (\bibinfo {year} {2018})}\BibitemShut {NoStop}%
\bibitem [{\citenamefont {Yapparov}\ \emph {et~al.}(2020)\citenamefont
  {Yapparov}, \citenamefont {Chow}, \citenamefont {Lynsky}, \citenamefont {Wu},
  \citenamefont {Nakamura}, \citenamefont {Speck},\ and\ \citenamefont
  {Marcinkevi{\v{c}}ius}}]{Vdefect14}%
  \BibitemOpen
  \bibfield  {author} {\bibinfo {author} {\bibfnamefont {R.}~\bibnamefont
  {Yapparov}}, \bibinfo {author} {\bibfnamefont {Y.~C.}\ \bibnamefont {Chow}},
  \bibinfo {author} {\bibfnamefont {C.}~\bibnamefont {Lynsky}}, \bibinfo
  {author} {\bibfnamefont {F.}~\bibnamefont {Wu}}, \bibinfo {author}
  {\bibfnamefont {S.}~\bibnamefont {Nakamura}}, \bibinfo {author}
  {\bibfnamefont {J.~S.}\ \bibnamefont {Speck}}, \ and\ \bibinfo {author}
  {\bibfnamefont {S.}~\bibnamefont {Marcinkevi{\v{c}}ius}},\ }\bibfield
  {title} {\enquote {\bibinfo {title} {Variations of light emission and carrier
  dynamics around {V-defects in InGaN} quantum wells},}\ }\href@noop {}
  {\bibfield  {journal} {\bibinfo  {journal} {Journal of Applied Physics}\
  }\textbf {\bibinfo {volume} {128}},\ \bibinfo {pages} {225703} (\bibinfo
  {year} {2020})}\BibitemShut {NoStop}%
\bibitem [{\citenamefont {Islam}\ \emph {et~al.}(2022)\citenamefont {Islam},
  \citenamefont {Kim}, \citenamefont {Shin}, \citenamefont {Shim},\ and\
  \citenamefont {Kwak}}]{Vdefect20}%
  \BibitemOpen
  \bibfield  {author} {\bibinfo {author} {\bibfnamefont {A.~B. M.~H.}\
  \bibnamefont {Islam}}, \bibinfo {author} {\bibfnamefont {T.~K.}\ \bibnamefont
  {Kim}}, \bibinfo {author} {\bibfnamefont {D.-S.}\ \bibnamefont {Shin}},
  \bibinfo {author} {\bibfnamefont {J.-I.}\ \bibnamefont {Shim}}, \ and\
  \bibinfo {author} {\bibfnamefont {J.~S.}\ \bibnamefont {Kwak}},\ }\bibfield
  {title} {\enquote {\bibinfo {title} {Generation of sidewall defects in
  {InGaN/GaN} blue {micro-LEDs} under forward-current stress},}\ }\href@noop {}
  {\bibfield  {journal} {\bibinfo  {journal} {Applied Physics Letters}\
  }\textbf {\bibinfo {volume} {121}},\ \bibinfo {pages} {013501} (\bibinfo
  {year} {2022})}\BibitemShut {NoStop}%
\bibitem [{\citenamefont {Wu}\ \emph {et~al.}(2023)\citenamefont {Wu},
  \citenamefont {Ewing}, \citenamefont {Lynsky}, \citenamefont {Iza},
  \citenamefont {Nakamura}, \citenamefont {DenBaars},\ and\ \citenamefont
  {Speck}}]{Vdefect10}%
  \BibitemOpen
  \bibfield  {author} {\bibinfo {author} {\bibfnamefont {F.}~\bibnamefont
  {Wu}}, \bibinfo {author} {\bibfnamefont {J.}~\bibnamefont {Ewing}}, \bibinfo
  {author} {\bibfnamefont {C.}~\bibnamefont {Lynsky}}, \bibinfo {author}
  {\bibfnamefont {M.}~\bibnamefont {Iza}}, \bibinfo {author} {\bibfnamefont
  {S.}~\bibnamefont {Nakamura}}, \bibinfo {author} {\bibfnamefont {S.~P.}\
  \bibnamefont {DenBaars}}, \ and\ \bibinfo {author} {\bibfnamefont {J.~S.}\
  \bibnamefont {Speck}},\ }\bibfield  {title} {\enquote {\bibinfo {title}
  {Structure of {V-defects} in long wavelength {GaN}-based light emitting
  diodes},}\ }\href@noop {} {\bibfield  {journal} {\bibinfo  {journal} {Journal
  of Applied Physics}\ }\textbf {\bibinfo {volume} {133}},\ \bibinfo {pages}
  {035703} (\bibinfo {year} {2023})}\BibitemShut {NoStop}%
\bibitem [{\citenamefont {Marcinkevi{\v{c}}ius}\ \emph
  {et~al.}(2024)\citenamefont {Marcinkevi{\v{c}}ius}, \citenamefont {Tak},
  \citenamefont {Chow}, \citenamefont {Wu}, \citenamefont {Yapparov},
  \citenamefont {DenBaars}, \citenamefont {Nakamura},\ and\ \citenamefont
  {Speck}}]{Vdefect15}%
  \BibitemOpen
  \bibfield  {author} {\bibinfo {author} {\bibfnamefont {S.}~\bibnamefont
  {Marcinkevi{\v{c}}ius}}, \bibinfo {author} {\bibfnamefont {T.}~\bibnamefont
  {Tak}}, \bibinfo {author} {\bibfnamefont {Y.~C.}\ \bibnamefont {Chow}},
  \bibinfo {author} {\bibfnamefont {F.}~\bibnamefont {Wu}}, \bibinfo {author}
  {\bibfnamefont {R.}~\bibnamefont {Yapparov}}, \bibinfo {author}
  {\bibfnamefont {S.~P.}\ \bibnamefont {DenBaars}}, \bibinfo {author}
  {\bibfnamefont {S.}~\bibnamefont {Nakamura}}, \ and\ \bibinfo {author}
  {\bibfnamefont {J.~S.}\ \bibnamefont {Speck}},\ }\bibfield  {title} {\enquote
  {\bibinfo {title} {Dynamics of carrier injection through {V-defects} in long
  wavelength {GaN} {LEDs}},}\ }\href@noop {} {\bibfield  {journal} {\bibinfo
  {journal} {Applied Physics Letters}\ }\textbf {\bibinfo {volume} {124}},\
  \bibinfo {pages} {181108} (\bibinfo {year} {2024})}\BibitemShut {NoStop}%
\bibitem [{\citenamefont {Han}\ \emph {et~al.}(2013)\citenamefont {Han},
  \citenamefont {Lee}, \citenamefont {Shim}, \citenamefont {Wook~Lee},
  \citenamefont {Kim}, \citenamefont {Yoon}, \citenamefont {Sun~Kim},\ and\
  \citenamefont {Kim}}]{Vdefect6}%
  \BibitemOpen
  \bibfield  {author} {\bibinfo {author} {\bibfnamefont {S.-H.}\ \bibnamefont
  {Han}}, \bibinfo {author} {\bibfnamefont {D.-Y.}\ \bibnamefont {Lee}},
  \bibinfo {author} {\bibfnamefont {H.-W.}\ \bibnamefont {Shim}}, \bibinfo
  {author} {\bibfnamefont {J.}~\bibnamefont {Wook~Lee}}, \bibinfo {author}
  {\bibfnamefont {D.-J.}\ \bibnamefont {Kim}}, \bibinfo {author} {\bibfnamefont
  {S.}~\bibnamefont {Yoon}}, \bibinfo {author} {\bibfnamefont {Y.}~\bibnamefont
  {Sun~Kim}}, \ and\ \bibinfo {author} {\bibfnamefont {S.-T.}\ \bibnamefont
  {Kim}},\ }\bibfield  {title} {\enquote {\bibinfo {title} {Improvement of
  efficiency and electrical properties using intentionally formed {V-shaped}
  pits in {InGaN/GaN} multiple quantum well light-emitting diodes},}\
  }\href@noop {} {\bibfield  {journal} {\bibinfo  {journal} {Applied Physics
  Letters}\ }\textbf {\bibinfo {volume} {102}},\ \bibinfo {pages} {251123}
  (\bibinfo {year} {2013})}\BibitemShut {NoStop}%
\bibitem [{\citenamefont {Li}\ \emph {et~al.}(2016)\citenamefont {Li},
  \citenamefont {Wu}, \citenamefont {Hsu}, \citenamefont {Lu}, \citenamefont
  {Li}, \citenamefont {Lu},\ and\ \citenamefont {Wu}}]{Vdefect8}%
  \BibitemOpen
  \bibfield  {author} {\bibinfo {author} {\bibfnamefont {C.-K.}\ \bibnamefont
  {Li}}, \bibinfo {author} {\bibfnamefont {C.-K.}\ \bibnamefont {Wu}}, \bibinfo
  {author} {\bibfnamefont {C.-C.}\ \bibnamefont {Hsu}}, \bibinfo {author}
  {\bibfnamefont {L.-S.}\ \bibnamefont {Lu}}, \bibinfo {author} {\bibfnamefont
  {H.}~\bibnamefont {Li}}, \bibinfo {author} {\bibfnamefont {T.-C.}\
  \bibnamefont {Lu}}, \ and\ \bibinfo {author} {\bibfnamefont {Y.-R.}\
  \bibnamefont {Wu}},\ }\bibfield  {title} {\enquote {\bibinfo {title} {{3D}
  numerical modeling of the carrier transport and radiative efficiency for
  {InGaN/GaN} light emitting diodes with {V-shaped} pits},}\ }\href@noop {}
  {\bibfield  {journal} {\bibinfo  {journal} {AIP Advances}\ }\textbf {\bibinfo
  {volume} {6}},\ \bibinfo {pages} {055208} (\bibinfo {year}
  {2016})}\BibitemShut {NoStop}%
\bibitem [{\citenamefont {Hsiao}\ \emph {et~al.}(2023)\citenamefont {Hsiao},
  \citenamefont {Lee}, \citenamefont {Miao}, \citenamefont {Pai}, \citenamefont
  {Iida}, \citenamefont {Lin}, \citenamefont {Chen}, \citenamefont {Chow},
  \citenamefont {Lin}, \citenamefont {Horng} \emph {et~al.}}]{leakage6}%
  \BibitemOpen
  \bibfield  {author} {\bibinfo {author} {\bibfnamefont {F.-H.}\ \bibnamefont
  {Hsiao}}, \bibinfo {author} {\bibfnamefont {T.-Y.}\ \bibnamefont {Lee}},
  \bibinfo {author} {\bibfnamefont {W.-C.}\ \bibnamefont {Miao}}, \bibinfo
  {author} {\bibfnamefont {Y.-H.}\ \bibnamefont {Pai}}, \bibinfo {author}
  {\bibfnamefont {D.}~\bibnamefont {Iida}}, \bibinfo {author} {\bibfnamefont
  {C.-L.}\ \bibnamefont {Lin}}, \bibinfo {author} {\bibfnamefont {F.-C.}\
  \bibnamefont {Chen}}, \bibinfo {author} {\bibfnamefont {C.-W.}\ \bibnamefont
  {Chow}}, \bibinfo {author} {\bibfnamefont {C.-C.}\ \bibnamefont {Lin}},
  \bibinfo {author} {\bibfnamefont {R.-H.}\ \bibnamefont {Horng}},  \emph
  {et~al.},\ }\bibfield  {title} {\enquote {\bibinfo {title} {Investigations on
  the high performance of {InGaN} red micro-{LEDs} with single quantum well for
  visible light communication applications},}\ }\href@noop {} {\bibfield
  {journal} {\bibinfo  {journal} {Discover Nano}\ }\textbf {\bibinfo {volume}
  {18}},\ \bibinfo {pages} {95} (\bibinfo {year} {2023})}\BibitemShut {NoStop}%
\bibitem [{\citenamefont {Park}\ \emph {et~al.}(2024)\citenamefont {Park},
  \citenamefont {Youn}, \citenamefont {Baek}, \citenamefont {Chu},
  \citenamefont {Kim}, \citenamefont {Geum}, \citenamefont {Kim}, \citenamefont
  {Kim}, \citenamefont {Kuk}, \citenamefont {Park} \emph {et~al.}}]{leakage}%
  \BibitemOpen
  \bibfield  {author} {\bibinfo {author} {\bibfnamefont {J.}~\bibnamefont
  {Park}}, \bibinfo {author} {\bibfnamefont {E.-J.}\ \bibnamefont {Youn}},
  \bibinfo {author} {\bibfnamefont {W.~J.}\ \bibnamefont {Baek}}, \bibinfo
  {author} {\bibfnamefont {E.-K.}\ \bibnamefont {Chu}}, \bibinfo {author}
  {\bibfnamefont {H.~S.}\ \bibnamefont {Kim}}, \bibinfo {author} {\bibfnamefont
  {D.-M.}\ \bibnamefont {Geum}}, \bibinfo {author} {\bibfnamefont {J.~P.}\
  \bibnamefont {Kim}}, \bibinfo {author} {\bibfnamefont {B.~H.}\ \bibnamefont
  {Kim}}, \bibinfo {author} {\bibfnamefont {S.-H.}\ \bibnamefont {Kuk}},
  \bibinfo {author} {\bibfnamefont {H.-H.}\ \bibnamefont {Park}},  \emph
  {et~al.},\ }\bibfield  {title} {\enquote {\bibinfo {title} {Size-dependent
  optoelectronic characteristics of {InGaN/GaN} red {micro-LEDs} on 4-inch {Si}
  substrates: high pixel density arrays demonstration},}\ }\href@noop {}
  {\bibfield  {journal} {\bibinfo  {journal} {Optics Express}\ }\textbf
  {\bibinfo {volume} {32}},\ \bibinfo {pages} {24242--24250} (\bibinfo {year}
  {2024})}\BibitemShut {NoStop}%
\bibitem [{\citenamefont {Shen}\ \emph {et~al.}(2021)\citenamefont {Shen},
  \citenamefont {Weisbuch}, \citenamefont {Speck},\ and\ \citenamefont
  {Wu}}]{randomalloyfluctuation3}%
  \BibitemOpen
  \bibfield  {author} {\bibinfo {author} {\bibfnamefont {H.-T.}\ \bibnamefont
  {Shen}}, \bibinfo {author} {\bibfnamefont {C.}~\bibnamefont {Weisbuch}},
  \bibinfo {author} {\bibfnamefont {J.~S.}\ \bibnamefont {Speck}}, \ and\
  \bibinfo {author} {\bibfnamefont {Y.-R.}\ \bibnamefont {Wu}},\ }\bibfield
  {title} {\enquote {\bibinfo {title} {Three-dimensional modeling of
  minority-carrier lateral diffusion length including random alloy fluctuations
  in {(In, Ga) N and (Al, Ga) N} single quantum wells},}\ }\href@noop {}
  {\bibfield  {journal} {\bibinfo  {journal} {Physical Review Applied}\
  }\textbf {\bibinfo {volume} {16}},\ \bibinfo {pages} {024054} (\bibinfo
  {year} {2021})}\BibitemShut {NoStop}%
\bibitem [{\citenamefont {Der~Maur}\ \emph {et~al.}(2014)\citenamefont
  {Der~Maur}, \citenamefont {Barettin}, \citenamefont {Pecchia}, \citenamefont
  {Sacconi},\ and\ \citenamefont {Di~Carlo}}]{QCSE}%
  \BibitemOpen
  \bibfield  {author} {\bibinfo {author} {\bibfnamefont {M.~A.}\ \bibnamefont
  {Der~Maur}}, \bibinfo {author} {\bibfnamefont {D.}~\bibnamefont {Barettin}},
  \bibinfo {author} {\bibfnamefont {A.}~\bibnamefont {Pecchia}}, \bibinfo
  {author} {\bibfnamefont {F.}~\bibnamefont {Sacconi}}, \ and\ \bibinfo
  {author} {\bibfnamefont {A.}~\bibnamefont {Di~Carlo}},\ }\bibfield  {title}
  {\enquote {\bibinfo {title} {Effect of alloy fluctuations in {InGaN/GaN}
  quantum wells on optical emission strength},}\ }in\ \href@noop {} {\emph
  {\bibinfo {booktitle} {Numerical Simulation of Optoelectronic Devices,
  2014}}}\ (\bibinfo {organization} {IEEE},\ \bibinfo {year} {2014})\ pp.\
  \bibinfo {pages} {11--12}\BibitemShut {NoStop}%
\bibitem [{\citenamefont {Robertson}\ \emph {et~al.}(2019)\citenamefont
  {Robertson}, \citenamefont {Qwah}, \citenamefont {Wu},\ and\ \citenamefont
  {Speck}}]{TD3}%
  \BibitemOpen
  \bibfield  {author} {\bibinfo {author} {\bibfnamefont {C.}~\bibnamefont
  {Robertson}}, \bibinfo {author} {\bibfnamefont {K.}~\bibnamefont {Qwah}},
  \bibinfo {author} {\bibfnamefont {Y.-R.}\ \bibnamefont {Wu}}, \ and\ \bibinfo
  {author} {\bibfnamefont {J.}~\bibnamefont {Speck}},\ }\bibfield  {title}
  {\enquote {\bibinfo {title} {Modeling dislocation-related leakage currents in
  {GaN} pn diodes},}\ }\href@noop {} {\bibfield  {journal} {\bibinfo  {journal}
  {Journal of Applied Physics}\ }\textbf {\bibinfo {volume} {126}},\ \bibinfo
  {pages} {245705} (\bibinfo {year} {2019})}\BibitemShut {NoStop}%
\bibitem [{\citenamefont {Wu}\ \emph {et~al.}(2012)\citenamefont {Wu},
  \citenamefont {Shivaraman}, \citenamefont {Wang},\ and\ \citenamefont
  {Speck}}]{randomalloyfluctuation2}%
  \BibitemOpen
  \bibfield  {author} {\bibinfo {author} {\bibfnamefont {Y.-R.}\ \bibnamefont
  {Wu}}, \bibinfo {author} {\bibfnamefont {R.}~\bibnamefont {Shivaraman}},
  \bibinfo {author} {\bibfnamefont {K.-C.}\ \bibnamefont {Wang}}, \ and\
  \bibinfo {author} {\bibfnamefont {J.~S.}\ \bibnamefont {Speck}},\ }\bibfield
  {title} {\enquote {\bibinfo {title} {Analyzing the physical properties of
  {InGaN} multiple quantum well light emitting diodes from nano scale
  structure},}\ }\href@noop {} {\bibfield  {journal} {\bibinfo  {journal}
  {Applied Physics Letters}\ }\textbf {\bibinfo {volume} {101}},\ \bibinfo
  {pages} {083505} (\bibinfo {year} {2012})}\BibitemShut {NoStop}%
\bibitem [{\citenamefont {Filoche}\ \emph {et~al.}(2017)\citenamefont
  {Filoche}, \citenamefont {Piccardo}, \citenamefont {Wu}, \citenamefont {Li},
  \citenamefont {Weisbuch},\ and\ \citenamefont {Mayboroda}}]{LL1}%
  \BibitemOpen
  \bibfield  {author} {\bibinfo {author} {\bibfnamefont {M.}~\bibnamefont
  {Filoche}}, \bibinfo {author} {\bibfnamefont {M.}~\bibnamefont {Piccardo}},
  \bibinfo {author} {\bibfnamefont {Y.-R.}\ \bibnamefont {Wu}}, \bibinfo
  {author} {\bibfnamefont {C.-K.}\ \bibnamefont {Li}}, \bibinfo {author}
  {\bibfnamefont {C.}~\bibnamefont {Weisbuch}}, \ and\ \bibinfo {author}
  {\bibfnamefont {S.}~\bibnamefont {Mayboroda}},\ }\bibfield  {title} {\enquote
  {\bibinfo {title} {Localization landscape theory of disorder in
  semiconductors. {I.} {Theory} and modeling},}\ }\href@noop {} {\bibfield
  {journal} {\bibinfo  {journal} {Physical Review B}\ }\textbf {\bibinfo
  {volume} {95}},\ \bibinfo {pages} {144204} (\bibinfo {year}
  {2017})}\BibitemShut {NoStop}%
\bibitem [{\citenamefont {Piccardo}\ \emph {et~al.}(2017)\citenamefont
  {Piccardo}, \citenamefont {Li}, \citenamefont {Wu}, \citenamefont {Speck},
  \citenamefont {Bonef}, \citenamefont {Farrell}, \citenamefont {Filoche},
  \citenamefont {Martinelli}, \citenamefont {Peretti},\ and\ \citenamefont
  {Weisbuch}}]{LL3}%
  \BibitemOpen
  \bibfield  {author} {\bibinfo {author} {\bibfnamefont {M.}~\bibnamefont
  {Piccardo}}, \bibinfo {author} {\bibfnamefont {C.-K.}\ \bibnamefont {Li}},
  \bibinfo {author} {\bibfnamefont {Y.-R.}\ \bibnamefont {Wu}}, \bibinfo
  {author} {\bibfnamefont {J.~S.}\ \bibnamefont {Speck}}, \bibinfo {author}
  {\bibfnamefont {B.}~\bibnamefont {Bonef}}, \bibinfo {author} {\bibfnamefont
  {R.~M.}\ \bibnamefont {Farrell}}, \bibinfo {author} {\bibfnamefont
  {M.}~\bibnamefont {Filoche}}, \bibinfo {author} {\bibfnamefont
  {L.}~\bibnamefont {Martinelli}}, \bibinfo {author} {\bibfnamefont
  {J.}~\bibnamefont {Peretti}}, \ and\ \bibinfo {author} {\bibfnamefont
  {C.}~\bibnamefont {Weisbuch}},\ }\bibfield  {title} {\enquote {\bibinfo
  {title} {Localization landscape theory of disorder in semiconductors. {II.}
  {Urbach} tails of disordered quantum well layers},}\ }\href@noop {}
  {\bibfield  {journal} {\bibinfo  {journal} {Physical Review B}\ }\textbf
  {\bibinfo {volume} {95}},\ \bibinfo {pages} {144205} (\bibinfo {year}
  {2017})}\BibitemShut {NoStop}%
\bibitem [{\citenamefont {Casey~Jr}\ and\ \citenamefont {Stern}(1976)}]{tail8}%
  \BibitemOpen
  \bibfield  {author} {\bibinfo {author} {\bibfnamefont {H.}~\bibnamefont
  {Casey~Jr}}\ and\ \bibinfo {author} {\bibfnamefont {F.}~\bibnamefont
  {Stern}},\ }\bibfield  {title} {\enquote {\bibinfo {title}
  {Concentration-dependent absorption and spontaneous emission of heavily doped
  {GaAs}},}\ }\href@noop {} {\bibfield  {journal} {\bibinfo  {journal} {Journal
  of Applied Physics}\ }\textbf {\bibinfo {volume} {47}},\ \bibinfo {pages}
  {631--643} (\bibinfo {year} {1976})}\BibitemShut {NoStop}%
\bibitem [{\citenamefont {Kirchartz}\ \emph {et~al.}(2011)\citenamefont
  {Kirchartz}, \citenamefont {Pieters}, \citenamefont {Kirkpatrick},
  \citenamefont {Rau},\ and\ \citenamefont {Nelson}}]{tail4}%
  \BibitemOpen
  \bibfield  {author} {\bibinfo {author} {\bibfnamefont {T.}~\bibnamefont
  {Kirchartz}}, \bibinfo {author} {\bibfnamefont {B.~E.}\ \bibnamefont
  {Pieters}}, \bibinfo {author} {\bibfnamefont {J.}~\bibnamefont
  {Kirkpatrick}}, \bibinfo {author} {\bibfnamefont {U.}~\bibnamefont {Rau}}, \
  and\ \bibinfo {author} {\bibfnamefont {J.}~\bibnamefont {Nelson}},\
  }\bibfield  {title} {\enquote {\bibinfo {title} {Recombination via tail
  states in polythiophene: fullerene solar cells},}\ }\href@noop {} {\bibfield
  {journal} {\bibinfo  {journal} {Physical Review B—Condensed Matter and
  Materials Physics}\ }\textbf {\bibinfo {volume} {83}},\ \bibinfo {pages}
  {115209} (\bibinfo {year} {2011})}\BibitemShut {NoStop}%
\bibitem [{\citenamefont {Kung}\ \emph {et~al.}(2017)\citenamefont {Kung},
  \citenamefont {Huang}, \citenamefont {Huang}, \citenamefont {Tseng},
  \citenamefont {Leung}, \citenamefont {Chiu}, \citenamefont {Lee},\ and\
  \citenamefont {Wu}}]{tail1}%
  \BibitemOpen
  \bibfield  {author} {\bibinfo {author} {\bibfnamefont {T.-J.}\ \bibnamefont
  {Kung}}, \bibinfo {author} {\bibfnamefont {J.-Y.}\ \bibnamefont {Huang}},
  \bibinfo {author} {\bibfnamefont {J.-J.}\ \bibnamefont {Huang}}, \bibinfo
  {author} {\bibfnamefont {S.~H.}\ \bibnamefont {Tseng}}, \bibinfo {author}
  {\bibfnamefont {M.-K.}\ \bibnamefont {Leung}}, \bibinfo {author}
  {\bibfnamefont {T.-L.}\ \bibnamefont {Chiu}}, \bibinfo {author}
  {\bibfnamefont {J.-H.}\ \bibnamefont {Lee}}, \ and\ \bibinfo {author}
  {\bibfnamefont {Y.-R.}\ \bibnamefont {Wu}},\ }\bibfield  {title} {\enquote
  {\bibinfo {title} {Modeling of carrier transport in organic light emitting
  diode with random dopant effects by two-dimensional simulation},}\
  }\href@noop {} {\bibfield  {journal} {\bibinfo  {journal} {Optics Express}\
  }\textbf {\bibinfo {volume} {25}},\ \bibinfo {pages} {25492--25503} (\bibinfo
  {year} {2017})}\BibitemShut {NoStop}%
\bibitem [{\citenamefont {Huang}\ \emph {et~al.}(2020)\citenamefont {Huang},
  \citenamefont {Lee}, \citenamefont {Wu}, \citenamefont {Chen}, \citenamefont
  {Chiu}, \citenamefont {Huang}, \citenamefont {Leung},\ and\ \citenamefont
  {Chiu}}]{tail3}%
  \BibitemOpen
  \bibfield  {author} {\bibinfo {author} {\bibfnamefont {J.-Y.}\ \bibnamefont
  {Huang}}, \bibinfo {author} {\bibfnamefont {J.-H.}\ \bibnamefont {Lee}},
  \bibinfo {author} {\bibfnamefont {Y.-R.}\ \bibnamefont {Wu}}, \bibinfo
  {author} {\bibfnamefont {T.-Y.}\ \bibnamefont {Chen}}, \bibinfo {author}
  {\bibfnamefont {Y.-C.}\ \bibnamefont {Chiu}}, \bibinfo {author}
  {\bibfnamefont {J.-J.}\ \bibnamefont {Huang}}, \bibinfo {author}
  {\bibfnamefont {M.-k.}\ \bibnamefont {Leung}}, \ and\ \bibinfo {author}
  {\bibfnamefont {T.-L.}\ \bibnamefont {Chiu}},\ }\bibfield  {title} {\enquote
  {\bibinfo {title} {Revealing the mechanism of carrier transport in host-guest
  systems of organic materials with a modified {Poisson} and drift-diffusion
  solver},}\ }\href@noop {} {\bibfield  {journal} {\bibinfo  {journal}
  {Physical Review Materials}\ }\textbf {\bibinfo {volume} {4}},\ \bibinfo
  {pages} {125602} (\bibinfo {year} {2020})}\BibitemShut {NoStop}%
\bibitem [{\citenamefont {Huang}\ \emph {et~al.}(2023)\citenamefont {Huang},
  \citenamefont {Hung}, \citenamefont {Hsu}, \citenamefont {Chen},
  \citenamefont {Lee}, \citenamefont {Lin}, \citenamefont {Lin}, \citenamefont
  {Leung}, \citenamefont {Chiu}, \citenamefont {Lee} \emph {et~al.}}]{tail2}%
  \BibitemOpen
  \bibfield  {author} {\bibinfo {author} {\bibfnamefont {J.-Y.}\ \bibnamefont
  {Huang}}, \bibinfo {author} {\bibfnamefont {H.-C.}\ \bibnamefont {Hung}},
  \bibinfo {author} {\bibfnamefont {K.-C.}\ \bibnamefont {Hsu}}, \bibinfo
  {author} {\bibfnamefont {C.-H.}\ \bibnamefont {Chen}}, \bibinfo {author}
  {\bibfnamefont {P.-H.}\ \bibnamefont {Lee}}, \bibinfo {author} {\bibfnamefont
  {H.-Y.}\ \bibnamefont {Lin}}, \bibinfo {author} {\bibfnamefont {B.-Y.}\
  \bibnamefont {Lin}}, \bibinfo {author} {\bibfnamefont {M.-k.}\ \bibnamefont
  {Leung}}, \bibinfo {author} {\bibfnamefont {T.-L.}\ \bibnamefont {Chiu}},
  \bibinfo {author} {\bibfnamefont {J.-H.}\ \bibnamefont {Lee}},  \emph
  {et~al.},\ }\bibfield  {title} {\enquote {\bibinfo {title} {Numerical
  {Analysis} and {Optimization} of a {Hybrid} {Layer} {Structure} for
  {Triplet--Triplet} {Fusion} {Mechanism} in {Organic} {Light-Emitting}
  {Diodes}},}\ }\href@noop {} {\bibfield  {journal} {\bibinfo  {journal}
  {Advanced Theory and Simulations}\ }\textbf {\bibinfo {volume} {6}},\
  \bibinfo {pages} {2200633} (\bibinfo {year} {2023})}\BibitemShut {NoStop}%
\bibitem [{\citenamefont {Shahmohammadi}\ \emph {et~al.}(2017)\citenamefont
  {Shahmohammadi}, \citenamefont {Liu}, \citenamefont {Rossbach}, \citenamefont
  {Lahourcade}, \citenamefont {Dussaigne}, \citenamefont {Bougerol},
  \citenamefont {Butt{\'e}}, \citenamefont {Grandjean}, \citenamefont
  {Deveaud},\ and\ \citenamefont {Jacopin}}]{Auger}%
  \BibitemOpen
  \bibfield  {author} {\bibinfo {author} {\bibfnamefont {M.}~\bibnamefont
  {Shahmohammadi}}, \bibinfo {author} {\bibfnamefont {W.}~\bibnamefont {Liu}},
  \bibinfo {author} {\bibfnamefont {G.}~\bibnamefont {Rossbach}}, \bibinfo
  {author} {\bibfnamefont {L.}~\bibnamefont {Lahourcade}}, \bibinfo {author}
  {\bibfnamefont {A.}~\bibnamefont {Dussaigne}}, \bibinfo {author}
  {\bibfnamefont {C.}~\bibnamefont {Bougerol}}, \bibinfo {author}
  {\bibfnamefont {R.}~\bibnamefont {Butt{\'e}}}, \bibinfo {author}
  {\bibfnamefont {N.}~\bibnamefont {Grandjean}}, \bibinfo {author}
  {\bibfnamefont {B.}~\bibnamefont {Deveaud}}, \ and\ \bibinfo {author}
  {\bibfnamefont {G.}~\bibnamefont {Jacopin}},\ }\bibfield  {title} {\enquote
  {\bibinfo {title} {Enhancement of {Auger} recombination induced by carrier
  localization in {InGaN/GaN} quantum wells},}\ }\href@noop {} {\bibfield
  {journal} {\bibinfo  {journal} {Physical Review B}\ }\textbf {\bibinfo
  {volume} {95}},\ \bibinfo {pages} {125314} (\bibinfo {year}
  {2017})}\BibitemShut {NoStop}%
\bibitem [{\citenamefont {Ryu}\ \emph {et~al.}(2020)\citenamefont {Ryu},
  \citenamefont {Ryu}, \citenamefont {Onwukaeme},\ and\ \citenamefont
  {Ma}}]{Auger1}%
  \BibitemOpen
  \bibfield  {author} {\bibinfo {author} {\bibfnamefont {H.-Y.}\ \bibnamefont
  {Ryu}}, \bibinfo {author} {\bibfnamefont {G.-H.}\ \bibnamefont {Ryu}},
  \bibinfo {author} {\bibfnamefont {C.}~\bibnamefont {Onwukaeme}}, \ and\
  \bibinfo {author} {\bibfnamefont {B.}~\bibnamefont {Ma}},\ }\bibfield
  {title} {\enquote {\bibinfo {title} {Temperature dependence of the {Auger}
  recombination coefficient in {InGaN/GaN} multiple-quantum-well light-emitting
  diodes},}\ }\href@noop {} {\bibfield  {journal} {\bibinfo  {journal} {Optics
  Express}\ }\textbf {\bibinfo {volume} {28}},\ \bibinfo {pages} {27459--27472}
  (\bibinfo {year} {2020})}\BibitemShut {NoStop}%
\bibitem [{\citenamefont {Duboz}\ \emph {et~al.}(2023)\citenamefont {Duboz},
  \citenamefont {Isnard}, \citenamefont {Zuniga-Perez},\ and\ \citenamefont
  {Massies}}]{DUBOZ2023127033}%
  \BibitemOpen
  \bibfield  {author} {\bibinfo {author} {\bibfnamefont {J.-Y.}\ \bibnamefont
  {Duboz}}, \bibinfo {author} {\bibfnamefont {W.}~\bibnamefont {Isnard}},
  \bibinfo {author} {\bibfnamefont {J.}~\bibnamefont {Zuniga-Perez}}, \ and\
  \bibinfo {author} {\bibfnamefont {J.}~\bibnamefont {Massies}},\ }\bibfield
  {title} {\enquote {\bibinfo {title} {Why and how {In} composition
  fluctuations appear in {InGaN}?}}\ }\href {\doibase
  https://doi.org/10.1016/j.jcrysgro.2022.127033} {\bibfield  {journal}
  {\bibinfo  {journal} {Journal of Crystal Growth}\ }\textbf {\bibinfo {volume}
  {603}},\ \bibinfo {pages} {127033} (\bibinfo {year} {2023})}\BibitemShut
  {NoStop}%
\bibitem [{\citenamefont {Liu}\ \emph {et~al.}(2014)\citenamefont {Liu},
  \citenamefont {Huang}, \citenamefont {Davis}, \citenamefont {Porter},
  \citenamefont {Schreiber}, \citenamefont {Kuchibatla}, \citenamefont
  {Shutthanandan}, \citenamefont {Thevuthasan}, \citenamefont {Preble},
  \citenamefont {Paskova},\ and\ \citenamefont {Evans}}]{10.1116}%
  \BibitemOpen
  \bibfield  {author} {\bibinfo {author} {\bibfnamefont {F.}~\bibnamefont
  {Liu}}, \bibinfo {author} {\bibfnamefont {L.}~\bibnamefont {Huang}}, \bibinfo
  {author} {\bibfnamefont {R.~F.}\ \bibnamefont {Davis}}, \bibinfo {author}
  {\bibfnamefont {L.~M.}\ \bibnamefont {Porter}}, \bibinfo {author}
  {\bibfnamefont {D.~K.}\ \bibnamefont {Schreiber}}, \bibinfo {author}
  {\bibfnamefont {S.~V. N.~T.}\ \bibnamefont {Kuchibatla}}, \bibinfo {author}
  {\bibfnamefont {V.}~\bibnamefont {Shutthanandan}}, \bibinfo {author}
  {\bibfnamefont {S.}~\bibnamefont {Thevuthasan}}, \bibinfo {author}
  {\bibfnamefont {E.~A.}\ \bibnamefont {Preble}}, \bibinfo {author}
  {\bibfnamefont {T.}~\bibnamefont {Paskova}}, \ and\ \bibinfo {author}
  {\bibfnamefont {K.~R.}\ \bibnamefont {Evans}},\ }\bibfield  {title} {\enquote
  {\bibinfo {title} {Composition and interface analysis of {InGaN/GaN}
  multiquantum-wells on {GaN} substrates using atom probe tomography},}\ }\href
  {\doibase 10.1116/1.4893976} {\bibfield  {journal} {\bibinfo  {journal}
  {Journal of Vacuum Science \& Technology B}\ }\textbf {\bibinfo {volume}
  {32}},\ \bibinfo {pages} {051209} (\bibinfo {year} {2014})}\BibitemShut
  {NoStop}%
\bibitem [{\citenamefont {Tsai}\ \emph {et~al.}(2023)\citenamefont {Tsai},
  \citenamefont {Qwah}, \citenamefont {Banon}, \citenamefont {Filoche},
  \citenamefont {Weisbuch}, \citenamefont {Wu},\ and\ \citenamefont
  {Speck}}]{indiumcomposition}%
  \BibitemOpen
  \bibfield  {author} {\bibinfo {author} {\bibfnamefont {T.-Y.}\ \bibnamefont
  {Tsai}}, \bibinfo {author} {\bibfnamefont {K.~S.}\ \bibnamefont {Qwah}},
  \bibinfo {author} {\bibfnamefont {J.-P.}\ \bibnamefont {Banon}}, \bibinfo
  {author} {\bibfnamefont {M.}~\bibnamefont {Filoche}}, \bibinfo {author}
  {\bibfnamefont {C.}~\bibnamefont {Weisbuch}}, \bibinfo {author}
  {\bibfnamefont {Y.-R.}\ \bibnamefont {Wu}}, \ and\ \bibinfo {author}
  {\bibfnamefont {J.~S.}\ \bibnamefont {Speck}},\ }\bibfield  {title} {\enquote
  {\bibinfo {title} {Carrier localization in {III-nitride} versus conventional
  {III-V} semiconductors: {A} study on the effects of alloy disorder using
  landscape theory and the {Schr{\"o}dinger} equation},}\ }\href@noop {}
  {\bibfield  {journal} {\bibinfo  {journal} {Physical Review Applied}\
  }\textbf {\bibinfo {volume} {20}},\ \bibinfo {pages} {044069} (\bibinfo
  {year} {2023})}\BibitemShut {NoStop}%
\bibitem [{\citenamefont {Huang}\ and\ \citenamefont {Wu}(2009)}]{tail7}%
  \BibitemOpen
  \bibfield  {author} {\bibinfo {author} {\bibfnamefont {H.-H.}\ \bibnamefont
  {Huang}}\ and\ \bibinfo {author} {\bibfnamefont {Y.-R.}\ \bibnamefont {Wu}},\
  }\bibfield  {title} {\enquote {\bibinfo {title} {Study of polarization
  properties of light emitted from a-plane {InGaN/GaN} quantum well-based light
  emitting diodes},}\ }\href@noop {} {\bibfield  {journal} {\bibinfo  {journal}
  {Journal of Applied Physics}\ }\textbf {\bibinfo {volume} {106}},\ \bibinfo
  {pages} {023106} (\bibinfo {year} {2009})}\BibitemShut {NoStop}%
\bibitem [{\citenamefont {Hasegawa}\ \emph {et~al.}(2007)\citenamefont
  {Hasegawa}, \citenamefont {Kamimura}, \citenamefont {Edagawa},\ and\
  \citenamefont {Yonenaga}}]{tail6}%
  \BibitemOpen
  \bibfield  {author} {\bibinfo {author} {\bibfnamefont {H.}~\bibnamefont
  {Hasegawa}}, \bibinfo {author} {\bibfnamefont {Y.}~\bibnamefont {Kamimura}},
  \bibinfo {author} {\bibfnamefont {K.}~\bibnamefont {Edagawa}}, \ and\
  \bibinfo {author} {\bibfnamefont {I.}~\bibnamefont {Yonenaga}},\ }\bibfield
  {title} {\enquote {\bibinfo {title} {Dislocation-related optical absorption
  in plastically deformed {GaN}},}\ }\href@noop {} {\bibfield  {journal}
  {\bibinfo  {journal} {Journal of Applied Physics}\ }\textbf {\bibinfo
  {volume} {102}},\ \bibinfo {pages} {026103} (\bibinfo {year}
  {2007})}\BibitemShut {NoStop}%
\bibitem [{\citenamefont {Zhao}\ \emph {et~al.}(2020)\citenamefont {Zhao},
  \citenamefont {Huang}, \citenamefont {Bruckbauer}, \citenamefont {Shen},
  \citenamefont {Zhu}, \citenamefont {Fletcher}, \citenamefont {Feng},
  \citenamefont {Cai}, \citenamefont {Bai}, \citenamefont {Trager-Cowan} \emph
  {et~al.}}]{zhao2020influence}%
  \BibitemOpen
  \bibfield  {author} {\bibinfo {author} {\bibfnamefont {X.}~\bibnamefont
  {Zhao}}, \bibinfo {author} {\bibfnamefont {K.}~\bibnamefont {Huang}},
  \bibinfo {author} {\bibfnamefont {J.}~\bibnamefont {Bruckbauer}}, \bibinfo
  {author} {\bibfnamefont {S.}~\bibnamefont {Shen}}, \bibinfo {author}
  {\bibfnamefont {C.}~\bibnamefont {Zhu}}, \bibinfo {author} {\bibfnamefont
  {P.}~\bibnamefont {Fletcher}}, \bibinfo {author} {\bibfnamefont
  {P.}~\bibnamefont {Feng}}, \bibinfo {author} {\bibfnamefont {Y.}~\bibnamefont
  {Cai}}, \bibinfo {author} {\bibfnamefont {J.}~\bibnamefont {Bai}}, \bibinfo
  {author} {\bibfnamefont {C.}~\bibnamefont {Trager-Cowan}},  \emph {et~al.},\
  }\bibfield  {title} {\enquote {\bibinfo {title} {Influence of an {InGaN}
  superlattice pre-layer on the performance of semi-polar (11--22) green {LEDs}
  grown on silicon},}\ }\href@noop {} {\bibfield  {journal} {\bibinfo
  {journal} {Scientific Reports}\ }\textbf {\bibinfo {volume} {10}},\ \bibinfo
  {pages} {12650} (\bibinfo {year} {2020})}\BibitemShut {NoStop}%
\bibitem [{\citenamefont {Tao}\ \emph {et~al.}(2018)\citenamefont {Tao},
  \citenamefont {Liu}, \citenamefont {Zhang}, \citenamefont {Mo}, \citenamefont
  {Xu}, \citenamefont {Ding}, \citenamefont {Wang}, \citenamefont {Wang},
  \citenamefont {Wu}, \citenamefont {Quan}, \citenamefont {Pan}, \citenamefont
  {Fang},\ and\ \citenamefont {Jiang}}]{Tao:18}%
  \BibitemOpen
  \bibfield  {author} {\bibinfo {author} {\bibfnamefont {X.}~\bibnamefont
  {Tao}}, \bibinfo {author} {\bibfnamefont {J.}~\bibnamefont {Liu}}, \bibinfo
  {author} {\bibfnamefont {J.}~\bibnamefont {Zhang}}, \bibinfo {author}
  {\bibfnamefont {C.}~\bibnamefont {Mo}}, \bibinfo {author} {\bibfnamefont
  {L.}~\bibnamefont {Xu}}, \bibinfo {author} {\bibfnamefont {J.}~\bibnamefont
  {Ding}}, \bibinfo {author} {\bibfnamefont {G.}~\bibnamefont {Wang}}, \bibinfo
  {author} {\bibfnamefont {X.}~\bibnamefont {Wang}}, \bibinfo {author}
  {\bibfnamefont {X.}~\bibnamefont {Wu}}, \bibinfo {author} {\bibfnamefont
  {Z.}~\bibnamefont {Quan}}, \bibinfo {author} {\bibfnamefont {S.}~\bibnamefont
  {Pan}}, \bibinfo {author} {\bibfnamefont {F.}~\bibnamefont {Fang}}, \ and\
  \bibinfo {author} {\bibfnamefont {F.}~\bibnamefont {Jiang}},\ }\bibfield
  {title} {\enquote {\bibinfo {title} {Performance enhancement of yellow
  {InGaN}-based multiple-quantum-well light-emitting diodes grown on {Si}
  substrates by optimizing the {InGaN/GaN} superlattice interlayer},}\ }\href
  {\doibase 10.1364/OME.8.001221} {\bibfield  {journal} {\bibinfo  {journal}
  {Opt. Mater. Express}\ }\textbf {\bibinfo {volume} {8}},\ \bibinfo {pages}
  {1221--1230} (\bibinfo {year} {2018})}\BibitemShut {NoStop}%
\bibitem [{\citenamefont {Jiang}\ \emph {et~al.}(2019)\citenamefont {Jiang},
  \citenamefont {Zheng}, \citenamefont {Mo}, \citenamefont {Wang},
  \citenamefont {Zhang}, \citenamefont {Quan}, \citenamefont {Liu},\ and\
  \citenamefont {Jiang}}]{JIANG2019120}%
  \BibitemOpen
  \bibfield  {author} {\bibinfo {author} {\bibfnamefont {X.}~\bibnamefont
  {Jiang}}, \bibinfo {author} {\bibfnamefont {C.}~\bibnamefont {Zheng}},
  \bibinfo {author} {\bibfnamefont {C.}~\bibnamefont {Mo}}, \bibinfo {author}
  {\bibfnamefont {X.}~\bibnamefont {Wang}}, \bibinfo {author} {\bibfnamefont
  {J.}~\bibnamefont {Zhang}}, \bibinfo {author} {\bibfnamefont
  {Z.}~\bibnamefont {Quan}}, \bibinfo {author} {\bibfnamefont {J.}~\bibnamefont
  {Liu}}, \ and\ \bibinfo {author} {\bibfnamefont {F.}~\bibnamefont {Jiang}},\
  }\bibfield  {title} {\enquote {\bibinfo {title} {Interface modification of
  two-step grown {InGaN/GaN} superlattices preparing layers for {InGaN}-based
  green {LED} on silicon substrate},}\ }\href {\doibase
  https://doi.org/10.1016/j.spmi.2018.12.017} {\bibfield  {journal} {\bibinfo
  {journal} {Superlattices and Microstructures}\ }\textbf {\bibinfo {volume}
  {126}},\ \bibinfo {pages} {120--124} (\bibinfo {year} {2019})}\BibitemShut
  {NoStop}%
\bibitem [{\citenamefont {Wong}\ \emph {et~al.}(2019)\citenamefont {Wong},
  \citenamefont {Lee}, \citenamefont {Myers}, \citenamefont {Hwang},
  \citenamefont {Kearns}, \citenamefont {Li}, \citenamefont {Speck},
  \citenamefont {Nakamura},\ and\ \citenamefont {DenBaars}}]{leakage2}%
  \BibitemOpen
  \bibfield  {author} {\bibinfo {author} {\bibfnamefont {M.~S.}\ \bibnamefont
  {Wong}}, \bibinfo {author} {\bibfnamefont {C.}~\bibnamefont {Lee}}, \bibinfo
  {author} {\bibfnamefont {D.~J.}\ \bibnamefont {Myers}}, \bibinfo {author}
  {\bibfnamefont {D.}~\bibnamefont {Hwang}}, \bibinfo {author} {\bibfnamefont
  {J.~A.}\ \bibnamefont {Kearns}}, \bibinfo {author} {\bibfnamefont
  {T.}~\bibnamefont {Li}}, \bibinfo {author} {\bibfnamefont {J.~S.}\
  \bibnamefont {Speck}}, \bibinfo {author} {\bibfnamefont {S.}~\bibnamefont
  {Nakamura}}, \ and\ \bibinfo {author} {\bibfnamefont {S.~P.}\ \bibnamefont
  {DenBaars}},\ }\bibfield  {title} {\enquote {\bibinfo {title}
  {Size-independent peak efficiency of {III}-nitride
  micro-light-emitting-diodes using chemical treatment and sidewall
  passivation},}\ }\href@noop {} {\bibfield  {journal} {\bibinfo  {journal}
  {Applied Physics Express}\ }\textbf {\bibinfo {volume} {12}},\ \bibinfo
  {pages} {097004} (\bibinfo {year} {2019})}\BibitemShut {NoStop}%
\bibitem [{\citenamefont {Zhuang}, \citenamefont {Iida},\ and\ \citenamefont
  {Ohkawa}(2020)}]{leakage1}%
  \BibitemOpen
  \bibfield  {author} {\bibinfo {author} {\bibfnamefont {Z.}~\bibnamefont
  {Zhuang}}, \bibinfo {author} {\bibfnamefont {D.}~\bibnamefont {Iida}}, \ and\
  \bibinfo {author} {\bibfnamefont {K.}~\bibnamefont {Ohkawa}},\ }\bibfield
  {title} {\enquote {\bibinfo {title} {Effects of size on the electrical and
  optical properties of {InGaN-based} red light-emitting diodes},}\ }\href@noop
  {} {\bibfield  {journal} {\bibinfo  {journal} {Applied Physics Letters}\
  }\textbf {\bibinfo {volume} {116}},\ \bibinfo {pages} {173501} (\bibinfo
  {year} {2020})}\BibitemShut {NoStop}%
\bibitem [{\citenamefont {Zhuang}, \citenamefont {Iida},\ and\ \citenamefont
  {Ohkawa}(2021)}]{leakage3}%
  \BibitemOpen
  \bibfield  {author} {\bibinfo {author} {\bibfnamefont {Z.}~\bibnamefont
  {Zhuang}}, \bibinfo {author} {\bibfnamefont {D.}~\bibnamefont {Iida}}, \ and\
  \bibinfo {author} {\bibfnamefont {K.}~\bibnamefont {Ohkawa}},\ }\bibfield
  {title} {\enquote {\bibinfo {title} {Investigation of {InGaN}-based red/green
  micro-light-emitting diodes},}\ }\href@noop {} {\bibfield  {journal}
  {\bibinfo  {journal} {Optics Letters}\ }\textbf {\bibinfo {volume} {46}},\
  \bibinfo {pages} {1912--1915} (\bibinfo {year} {2021})}\BibitemShut {NoStop}%
\bibitem [{\citenamefont {Liu}\ \emph {et~al.}(2022)\citenamefont {Liu},
  \citenamefont {Sun}, \citenamefont {Malhotra}, \citenamefont {Pandey},
  \citenamefont {Wang}, \citenamefont {Wu}, \citenamefont {Sun},\ and\
  \citenamefont {Mi}}]{leakage4}%
  \BibitemOpen
  \bibfield  {author} {\bibinfo {author} {\bibfnamefont {X.}~\bibnamefont
  {Liu}}, \bibinfo {author} {\bibfnamefont {Y.}~\bibnamefont {Sun}}, \bibinfo
  {author} {\bibfnamefont {Y.}~\bibnamefont {Malhotra}}, \bibinfo {author}
  {\bibfnamefont {A.}~\bibnamefont {Pandey}}, \bibinfo {author} {\bibfnamefont
  {P.}~\bibnamefont {Wang}}, \bibinfo {author} {\bibfnamefont {Y.}~\bibnamefont
  {Wu}}, \bibinfo {author} {\bibfnamefont {K.}~\bibnamefont {Sun}}, \ and\
  \bibinfo {author} {\bibfnamefont {Z.}~\bibnamefont {Mi}},\ }\bibfield
  {title} {\enquote {\bibinfo {title} {{N}-polar {InGaN} nanowires: breaking
  the efficiency bottleneck of nano and micro {LEDs}},}\ }\href@noop {}
  {\bibfield  {journal} {\bibinfo  {journal} {Photonics Research}\ }\textbf
  {\bibinfo {volume} {10}},\ \bibinfo {pages} {587--593} (\bibinfo {year}
  {2022})}\BibitemShut {NoStop}%
\bibitem [{\citenamefont {Pandey}, \citenamefont {Reddeppa},\ and\
  \citenamefont {Mi}(2024)}]{leakage5}%
  \BibitemOpen
  \bibfield  {author} {\bibinfo {author} {\bibfnamefont {A.}~\bibnamefont
  {Pandey}}, \bibinfo {author} {\bibfnamefont {M.}~\bibnamefont {Reddeppa}}, \
  and\ \bibinfo {author} {\bibfnamefont {Z.}~\bibnamefont {Mi}},\ }\bibfield
  {title} {\enquote {\bibinfo {title} {Recent progress on {micro-LEDs}},}\
  }\href@noop {} {\bibfield  {journal} {\bibinfo  {journal} {Light: Advanced
  Manufacturing}\ }\textbf {\bibinfo {volume} {4}},\ \bibinfo {pages}
  {519--542} (\bibinfo {year} {2024})}\BibitemShut {NoStop}%
\end{thebibliography}%

\end{document}